\documentclass[widetext,showpacs,preprintnumbers,superscriptaddress,nofootinbib]{revtex4}

\usepackage[dvipdfm]{graphicx}
\usepackage{amssymb}
\usepackage{amsmath}
\usepackage{dcolumn}
\usepackage{bm}
\usepackage{natbib}
\usepackage{multirow}
\usepackage{subfigure}
\usepackage{color}
\input{colordvi.tex}

\makeatletter
\@addtoreset{equation}{section}

\makeatother

\begin{document}

\preprint{KEK-TH-1726,\ UT-14-21}

\title{Affleck-Dine baryogenesis with R-parity violation}

\author{Tetsutaro Higaki}
\email{thigaki@post.kek.jp}
\affiliation{Theory Center, High Energy Accelerator Research Organization (KEK),
1-1 Oho, Tsukuba, Ibaraki 305-0801, Japan}

\author{Kazunori Nakayama}
\email{kazunori@hep-th.phys.s.u-tokyo.ac.jp}
\affiliation{Department of Physics, Graduate School of Science, University of Tokyo,
7-3-1 Bunkyo-ku, Hongo, Tokyo 113-0033, Japan}
\affiliation{Kavli Institute for the Physics and Mathematics of the Universe (WPI), Todai Institutes for Advanced Study, The University of Tokyo,
5-1-5 Kashiwa-no-ha, Kashiwa City, Chiba 277-8582, Japan}

\author{Ken'ichi Saikawa}
\email{saikawa@th.phys.titech.ac.jp}
\affiliation{Department of Physics, Tokyo Institute of Technology,
2-12-1 Ookayama, Meguro-ku, Tokyo 152-8551, Japan}

\author{Tomo Takahashi}
\email{tomot@cc.saga-u.ac.jp}
\affiliation{Department of Physics, Saga University, Saga 840-8502, Japan}

\author{Masahide Yamaguchi}
\email{gucci@phys.titech.ac.jp}
\affiliation{Department of Physics, Tokyo Institute of Technology,
2-12-1 Ookayama, Meguro-ku, Tokyo 152-8551, Japan}
 
\date{\today}

\begin{abstract}
We investigate whether the baryon asymmetry of the universe is explained in the framework of the supersymmetric
extension of the Standard Model with R-parity violating interactions.
It is shown that the Affleck-Dine mechanism naturally works via a trilinear interaction $LLE^c$, $LQD^c$, or $U^cD^cD^c$,
if the magnitude of the coupling corresponding to the operator $\lambda$, $\lambda'$, or $\lambda''$ is sufficiently small.
The formation of Q-balls and their subsequent evolution are also discussed.
The present baryon asymmetry can be explained in the parameter region where
R-parity is mildly violated $10^{-9}\lesssim\lambda,\lambda',\lambda''\lesssim 10^{-6}$ and
the mass of the gravitino is relatively heavy $m_{3/2}\gtrsim 10^4\mathrm{GeV}$.
On the other hand, it is difficult to explain the present baryon asymmetry for larger values of R-parity violating couplings
$\lambda,\lambda',\lambda''\gtrsim 10^{-5}$, since Q-balls are likely to be destructed in the thermal environment
and the primordial baryon number is washed away.
\end{abstract}


\maketitle

\begin{widetext}
\tableofcontents\vspace{5mm}
\end{widetext}

%%%%%%%%%%%%%%%%%%%%%%%%%%%%%%%%%%%%%%%%%%%%%%%%%%%%%%%%%%%%%%%%
\section{\label{sec1}Introduction}
%%%%%%%%%%%%%%%%%%%%%%%%%%%%%%%%%%%%%%%%%%%%%%%%%%%%%%%%%%%%%%%%
The new particle discovered at the Large Hadron Collider (LHC)~\cite{Aad:2012tfa,Chatrchyan:2012ufa} has been confirmed to be consistent with the Higgs boson in the Standard Model (SM),
and this fact strengthens the validity of the SM as a fundamental theory of elementary particles.
On the other hand, it is presumed that there are various phenomena in the early universe
which cannot be addressed in the framework of the SM.
In particular, it is widely believed that at the beginning of the universe
there must be the stage of the accelerated expansion of the universe called inflation~\cite{Linde:2005ht,Lyth:1998xn}.
The occurrence of inflation solves several puzzles such as the flatness and horizon problems~\cite{Guth:1980zm},
but it dilutes away the primordial baryon (B) number, implying that the creation of the B number (baryogenesis) should occur
after inflation in order to explain tiny but non-vanishing ratio of baryons to photons $\eta\approx 5\times 10^{-10}$
required by the big-bang nucleosynthesis (BBN) and cosmic microwave background~\cite{Weinberg:2008zzc}.
This mystery motivates us to extend the SM to a more fundamental theory.
In this paper, we focus on supersymmetry as a possible extension of the SM.

In supersymmetric field theories, generically there exist so-called the flat directions,
in which the potential for scalar fields vanishes in the absence of the soft supersymmetry breaking effects and nonrenormalizable operators.
In the context of cosmology, it is expected that some scalar fields corresponding to such flat directions get large expectation values during/after inflation.
If such scalar fields have nonvanishing B and lepton (L) numbers with some B and/or L violating interactions, which is true for the supersymmetric extension of the SM,
the primordial B asymmetry can be generated via the dynamics of these scalar fields after inflation.
This kind of scenario for the generation of the B asymmetry of the universe is called the Affleck-Dine (AD) mechanism~\cite{Affleck:1984fy,Dine:1995kz} [for reviews, see Ref.~\cite{Dine:2003ax}].

In addition to this mechanism for baryogenesis, supersymmetry is attractive for various reasons, such as the grand unification of the fundamental forces
and the solution for the hierarchy problem [see e.g. Ref.~\cite{Martin:1997ns}].
However, the low scale supersymmetry as a solution to the hierarchy problem suffers from the tension with the recent results at the LHC
which indicates that no evidence for the new physics is observed at the weak scale.
This fact suggests that the minimal supersymmetric version of the SM (MSSM) becomes unnatural, and
the supersymmetric model of the SM should be extended beyond the minimal one.
One of the possible extensions is to introduce the violation of R-parity.

R-parity~\cite{Farrar:1978xj} is to impose the parity symmetry $R_p=(-1)^{2S+3B+L}$ for the field content of the MSSM, where $S$, $B$, and $L$ are the spin, B number, and L number
of the particle, respectively. This parity is well motivated from the phenomenological point of view, since it suppresses the unwanted proton decay (see Sec.~\ref{sec2-1}),
and ensures the stability of the lightest supersymmetric particle (LSP), which becomes a good candidate of dark matter.
However, there might be no fundamental reason to introduce the R-parity as a global symmetry, and in general it can be largely violated.
Indeed, sizable R-parity violating interactions can be obtained in the framework of grand unified theories~\cite{DiLuzio:2013ysa}.
The introduction of the R-parity violation is getting more attention these days,
since it modifies the signatures observed at the LHC, which relaxes the stringent limits on the masses of supersymmetric particles~\cite{Carpenter:2006hs,Allanach:2012vj,Bhattacherjee:2013gr,Graham:2014vya}.

The introduction of the R-parity violation has a lot of relevance to the scenario of baryogenesis,
since it gives B/L number violating operators in the renormalizable superpotential [see Eq.~\eqref{RPV_superpotential}].
In this paper, we argue that the AD mechanism for baryogenesis naturally works via these B/L violating interactions.
It is also known that the R-parity violating interactions, together with the sphaleron process~\cite{Kuzmin:1985mm} which remains in the thermal equilibrium until
the epoch of the electroweak phase transition (EWPT), erase the primordial B number~\cite{Bouquet:1986mq,Campbell:1990fa,Fischler:1990gn,Dreiner:1992vm,Campbell:1991at}.
We will see that the estimation of the final B asymmetry becomes nontrivial because of these two competing factors.
Regarding this fact, we can consider the following two possibilities:
One is the scenario where the baryogenesis occurs with a sufficiently small R-parity violating coupling such that the erasure effect becomes ineffective.
Another scenario is that the primordial B number is preserved in the form of the nontopological soliton called Q-balls~\cite{Enqvist:1997si}, which protect the B asymmetry of the universe
from the erasure effect until the epoch of the EWPT.
We will show that the later scenario suffers from the difficulty to explain the present B asymmetry
because the destruction effects of Q-balls become significant in the thermal environment.

It should be noted that LSP is not stable in the scenario considered in this paper.
Therefore, LSPs do not play a role of dark matter of the universe,
which should be explained by other candidates such as axions~\cite{Preskill:1982cy,Abbott:1982af,Dine:1982ah}.\footnote{See also Refs.~\cite{Higaki:2014ooa,Marsh:2014qoa,Visinelli:2014twa}
for an isocurvature constraint generated by the QCD axion dark matter after BICEP2~\cite{Ade:2014xna}.}
In particular, it is difficult to resolve the coincidence between abundances of baryon and dark matter~\cite{Enqvist:1998en}
in the framework of the model presented in this paper.

The organization of this paper is as follows.
In Sec.~\ref{sec2}, we summarize the current constraints on the R-parity violating couplings.
These include indirect bounds based on the results of experimental studies and cosmological bounds such as 
the dissociation of light elements and the erasure of the primordial B number.
In Sec.~\ref{sec3}, the AD mechanism by the use of the R-parity violating operators is discussed.
We solve the evolution of the scalar field condensate after inflation, and estimate the B number generated from this mechanism.
Then, we introduce the finite temperature effects and derive some conditions for Q-balls to survive against the destruction effects
such as the evaporation into the surrounding plasma and the instability caused by the $U(1)$ violating operator in Sec.~\ref{sec4}.
Combining these results, we investigate the parameter region where the baryogenesis occurs successfully in Sec.~\ref{sec5}.
Finally, we conclude in Sec.~\ref{sec6}.

%%%%%%%%%%%%%%%%%%%%%%%%%%%%%%%%%%%%%%%%%%%%%%%%%%%%%%%%%%%%%%%%
\section{\label{sec2}Constraints on R-parity violating interactions}
%%%%%%%%%%%%%%%%%%%%%%%%%%%%%%%%%%%%%%%%%%%%%%%%%%%%%%%%%%%%%%%%
The general renormalizable R-parity violating superpotential allowed by the gauge invariance and the field content of the MSSM can be written as
\begin{equation}
W_{\not{R_p}} = \mu_iH_u L_i + \frac{1}{2}\lambda_{ijk}L_iL_jE^c_k
+ \lambda'_{ijk}L_iQ_jD^c_k + \frac{1}{2}\lambda_{ijk}''U_i^cD_j^cD_k^c, \label{RPV_superpotential}
\end{equation}
where $i,j,k=1,2,3$ are family indices, and contraction over gauge indices is understood.
Because of the gauge invariant contractions of $SU(2)_L$ fields $L_iL_j$ and $SU(3)_C$ fields $U_i^cD_j^cD_k^c$,
the coupling $\lambda_{ijk}$ is antisymmetric in its first two indices, and $\lambda_{ijk}''$ is antisymmetric in its last two indices.
The terms in the superpotential [Eq.~\eqref{RPV_superpotential}] can be divided into either B number or L number violating interactions:
\begin{align}
W_{\Delta L=1} &= \mu_iH_u L_i + \frac{1}{2}\lambda_{ijk}L_iL_jE^c_k
+ \lambda'_{ijk}L_iQ_jD^c_k, \\
W_{\Delta B=1} &= \frac{1}{2}\lambda_{ijk}''U_i^cD_j^cD_k^c,
\end{align}
where quantum numbers are assigned such that $B=+1/3$ for $Q_i$, $B=-1/3$ for $U^c_i$ and $D^c_i$,
$L=+1$ for $L_i$, and $L=-1$ for $E^c_i$.
The existence of these B or L violating couplings is tightly constrained from various observations.
In this section, we briefly overview the constraints on the magnitude of R-parity violating couplings $\mu_i$, $\lambda_{ijk}$, $\lambda_{ijk}'$, and $\lambda_{ijk}''$.
It should be emphasized that there are in general 96 (complex) independent R-parity violating parameters, and we will not quote every bound on these parameters.
Instead, we just enumerate several constraints which are relevant to our discussions on cosmology.
A more comprehensive review is found in~\cite{Barbier:2004ez} [see also~\cite{Kao:2009fg} for a recent update on indirect bounds].

%%%%%%%%%%%%%%%%%%%%%%%%%%%%%%%%%%%%%%%%%%%%%%%%%%%%%%%%%%%%%%%%
\subsection{\label{sec2-1}Single nucleon decay}
%%%%%%%%%%%%%%%%%%%%%%%%%%%%%%%%%%%%%%%%%%%%%%%%%%%%%%%%%%%%%%%%
The most severe constraints on the trilinear R-parity violating couplings $\lambda_{ijk}$, $\lambda_{ijk}'$, and $\lambda_{ijk}''$ are obtained from 
observations of single nucleon decay~\cite{Hinchliffe:1992ad}.
The combination of two operators in the superpotential $\lambda'_{ijk}L_iQ_jD^c_k$ and $\lambda_{ijk}''U_i^cD_j^cD_k^c/2$
leads to decay processes such as $p\to \pi^0l^+$, $n\to\pi^0\bar{\nu}$, and $p\to K^+\bar{\nu}$ mediated by $\tilde{d}_R$ squark in s-channel.
From the experimental lower bound on nucleon lifetimes, an upper bound on the coupling products $\lambda_{imk}'\lambda_{11k}''^*$ with
$i,k=1,2,3,$ $m=1,2$ can be obtained as~\cite{Barbier:2004ez,Beringer:1900zz}
\begin{equation}
|\lambda_{imk}'\lambda_{11k}''^*| < \mathcal{O}(1) \times 10^{-27}\left(\frac{100\mathrm{GeV}}{m_{\tilde{d}}}\right)^2, \label{nucleon_decay}
\end{equation}
where $m_{\tilde{d}}$ is the mass of the intermediate state down type squark.
A similar bound is obtained for the products $\lambda_{l1k}'\lambda_{12k}''^*$ ($l=1,2,$ $k=1,2,3$) from $\tilde{d}_R$ squark t-channel exchange diagrams.
Furthermore, decay processes such as $p\to\pi^+(K^+)e^{\pm}\mu^{\mp}\bar{\nu}$, $p\to\pi^+(K^+)\nu\bar{\nu}\bar{\nu}$, and
$p\to\pi^+(K^+)\bar{\nu}$ can occur due to diagrams mediated by virtual neutralinos, which lead to bounds on the products $\lambda_{i'j'k'}\lambda_{ijk}''^*$~\cite{Bhattacharyya:1998bx}.
Numerical values of these bounds depend on the family indices $(i,j,k)$ and $(i',j',k')$, such that $|\lambda_{i'j'k'}\lambda_{112}''^*|\lesssim 10^{-21}-10^{-14}$ and
$|\lambda_{i'j'k'}\lambda_{ijk}''^*|\lesssim 10^{-12}-10^{-3}$ for $(i,j,k)\ne(1,1,2)$.

Note that the stringent bound [Eq.~\eqref{nucleon_decay}] is applied on the product of $\lambda'$ and $\lambda''$.
If we assume a universal value for the R-parity violating couplings ($\lambda\simeq\lambda'\simeq\lambda''$),
this bound leads to\footnote{Here and hereafter, we use the notation ``$\lambda$" to represent the typical magnitude of trilinear R-parity violating couplings.}
\begin{equation}
\lambda < \mathcal{O}(1)\times 10^{-14}\left(\frac{1\mathrm{TeV}}{m_{\tilde{d}}}\right). \label{nucleon_decay_universal}
\end{equation}

%%%%%%%%%%%%%%%%%%%%%%%%%%%%%%%%%%%%%%%%%%%%%%%%%%%%%%%%%%%%%%%%
\subsection{\label{sec2-2}Light element bound}
%%%%%%%%%%%%%%%%%%%%%%%%%%%%%%%%%%%%%%%%%%%%%%%%%%%%%%%%%%%%%%%%
If R-parity violating couplings are sufficiently large, LSPs become unstable and eventually decay into lighter degrees of freedom.
Their lifetime is given by~\cite{Campbell:1991at}
\begin{equation}
\tau_{\rm LSP} \simeq 10^{-4}\mathrm{sec}\left(\frac{\lambda}{10^{-6}}\right)^{-2}\left(\frac{m_{\rm LSP}}{20\mathrm{GeV}}\right)^{-5}\left(\frac{m_{\tilde{f}}}{200\mathrm{GeV}}\right)^4,
\end{equation}
where $m_{\rm LSP}$ and $m_{\tilde{f}}$ are the masses of the LSP and sfermions, respectively.
Since the decay process via R-parity violating interactions involves hadronic energy injection, the late decay of LSPs is tightly constrained from the requirement
that it must not lead to the dissociation of light elements created during the epoch of BBN~\cite{Ellis:1990nb,Kim:1998mu}.
According to the constraint on late-decaying particles obtained in Refs.~\cite{Kawasaki:2004yh,Kawasaki:2004qu},
here we put the upper limit on the lifetime of LSPs, $\tau_{\rm LSP}\lesssim 1\mathrm{sec}$ with the assumption
that LSPs are not dark matter and their abundance is fixed by the decoupling from the thermal bath.\footnote{The abundance of the thermally produced neutralino
LSPs depends on their composition (i.e. Bino-, Wino-, or Higgsino-like)~\cite{ArkaniHamed:2006mb}.
Accordingly, the constraint from the light elements might vary by several orders of magnitude.
For simplicity, we ignore this model dependence and just fix the abundance so that it corresponds to the present dark matter abundance:
$m_{\rm LSP}Y_{\rm LSP}\simeq 4\times10^{-10}\mathrm{GeV}$,
where $Y_{\rm LSP}=n_{\rm LSP}/s$, $n_{\rm LSP}$ is the number density of LSPs, and $s$ is the entropy density.
For this value on the relic abundance, the results of Refs.~\cite{Kawasaki:2004yh,Kawasaki:2004qu} indicates that the lifetime must be shorter than $\mathcal{O}(10^{-1}-10^2)\mathrm{sec}$.
Note that this upper bound on the lifetime also depends on the hadronic branching ratio.
Here we take $\tau_{\rm LSP}\lesssim 1\mathrm{sec}$ as a conservative bound.}
This condition leads to the \emph{lower} bound on the trilinear R-parity violating couplings:
\begin{equation}
\lambda > 4 \times 10^{-9}\left(\frac{m_{\rm LSP}}{100\mathrm{GeV}}\right)^{-5/2}\left(\frac{m_{\tilde{f}}}{1\mathrm{TeV}}\right)^2. \label{light_element_lambda}
\end{equation}
For a fixed value of $\lambda$, this condition becomes
\begin{equation}
m_{\tilde{f}} > 2\times 10^{-6}\mathrm{GeV}\left(\frac{m_{\tilde{f}}}{10m_{\rm LSP}}\right)^5\left(\frac{10^{-4}}{\lambda}\right)^2. \label{light_element_m}
\end{equation}

When we combine two constraints given by Eqs.~\eqref{nucleon_decay_universal} and~\eqref{light_element_lambda}, we see that the whole parameter region is excluded
for the case where two types of couplings are comparable $\lambda'\simeq\lambda''$.
On the other hand, the bound given by Eq.~\eqref{nucleon_decay_universal} collapses if we assume that either $\lambda'$ or $\lambda''$ is highly suppressed.
We adopt this assumption throughout this paper.

%%%%%%%%%%%%%%%%%%%%%%%%%%%%%%%%%%%%%%%%%%%%%%%%%%%%%%%%%%%%%%%%
\subsection{\label{sec2-3}Sphaleron erasure}
%%%%%%%%%%%%%%%%%%%%%%%%%%%%%%%%%%%%%%%%%%%%%%%%%%%%%%%%%%%%%%%%
As we mentioned in the Introduction, another stringent upper bound on the trilinear R-parity violating couplings is obtained from the requirement of baryogenesis.
The effect of the nonperturbative process mediated by field configurations known as sphalerons~\cite{Manton:1983nd,Klinkhamer:1984di}
becomes significant at temperatures higher than $\mathcal{O}(100)\mathrm{GeV}$,
which washes out the primordial (B$+$L) number but conserves the (B$-$L) number~\cite{Kuzmin:1985mm}.
In addition to this sphaleron process, if B and/or L violating interactions due to the R-parity violating operators are in the thermal equilibrium at the temperature above the weak scale,
they completely erase the primordial B/L asymmetry~\cite{Bouquet:1986mq,Campbell:1990fa,Fischler:1990gn,Dreiner:1992vm,Campbell:1991at}.
Here, we quote the bound obtained in Refs.~\cite{Dreiner:1992vm,Campbell:1991at}, which is based on the calculation of $2\to1$ processes between fermion pairs and sfermions.

The rate for $2\to1$ process due to the trilinear R-parity violating operator is given by~\cite{Dreiner:1992vm}
\begin{equation}
\Gamma = \frac{\lambda^2}{9\pi\zeta(3)}\frac{m_{\tilde{f}}^2}{T}f\bigg(\frac{m_{\tilde{f}}^2}{T^2}\bigg),
\end{equation}
where $\zeta(3)\simeq 1.202$ is Riemann zeta function and $f(x)$ is defined by
\begin{equation}
f(x) = \int^{\infty}_0\frac{\ln(1+e^{-\frac{x}{4y}})}{e^y+1}dy.
\end{equation}
Note that this process is relevant only at the temperature $T\gtrsim m_{\tilde{f}}$, since otherwise the initial particles do not have a sufficient energy to annihilate into sfermions.
The ratio between the annihilation rate and the expansion rate becomes
\begin{equation}
\frac{\Gamma}{H} = \frac{\sqrt{10}\lambda^2g_*(T)^{-1/2}}{3\pi^2\zeta(3)}\frac{m_{\tilde{f}}^2M_{\rm Pl}}{T^3}f\bigg(\frac{m_{\tilde{f}}^2}{T^2}\bigg),
\end{equation}
where $g_*(T)$ is the relativistic degrees of freedom at the temperature $T$.
Requiring that this process is out of equilibrium up to the temperature $T\gtrsim m_{\tilde{f}}$ ($\Gamma/H|_{T=m_{\tilde{f}}}<1$),
we obtain the upper bound on the R-parity violating couplings
\begin{equation}
\lambda < 4 \times 10^{-7}\left(\frac{g_*(m_{\tilde{f}})}{100}\right)^{1/4}\left(\frac{m_{\tilde{f}}}{1\mathrm{TeV}}\right)^{1/2}, \label{sphaleron_lambda}
\end{equation}
where we use $f(1)\simeq 0.3375$. For a fixed value of $\lambda$, this condition becomes a lower bound on $m_{\tilde{f}}$:
\begin{equation}
m_{\tilde{f}} > 7\times 10^4\mathrm{TeV}\left(\frac{g_*(m_{\tilde{f}})}{100}\right)^{-1/2}\left(\frac{\lambda}{10^{-4}}\right)^2 \label{sphaleron_m}.
\end{equation}
Note that this bound does not hold if the B number is generated 
below the temperature $T_{\rm erasure}$ at which the R-parity violating processes become out of equilibrium.\footnote{Since the rate for $2\to1$ process is exponentially suppressed for $T<m_{\tilde{f}}$,
we expect that the critical temperature below which the wash out effect becomes irrelevant is roughly estimated as $T_{\rm erasure}\approx\mathcal{O}(m_{\tilde{f}})$.}
We will discuss such a possibility in Secs.~\ref{sec4} and~\ref{sec5}.

%%%%%%%%%%%%%%%%%%%%%%%%%%%%%%%%%%%%%%%%%%%%%%%%%%%%%%%%%%%%%%%%
\subsection{\label{sec2-4}Constraints on bilinear R-parity violation}
%%%%%%%%%%%%%%%%%%%%%%%%%%%%%%%%%%%%%%%%%%%%%%%%%%%%%%%%%%%%%%%%
Next, let us comment on the magnitude of the bilinear R-parity violating couplings $\mu_i$.
We note that the term $\mu_iH_uL_i$ in the superpotential [Eq.~\eqref{RPV_superpotential}] can be rotated away
due to the redefinition of four doublet superfields $(L_i,H_d)$, if the soft supersymmetry breaking terms are absent.
On the other hand, when the soft supersymmetry breaking effects are included, it becomes impossible to eliminate $\mu_i$-terms
and the bilinear R-parity violating terms arising from soft terms simultaneously, and the description of the R-parity violating effects
depends on the choice of the basis of $(L_i,H_d)$.

Constraints on the bilinear R-parity violating effects can be parametrized in terms of two basis-independent quantities, $\sin\xi$ and $\sin\zeta$~\cite{Barbier:2004ez}.
The quantity $\sin\xi$ represents the effect of the bilinear R-parity violation in the fermion sector~\cite{Banks:1995by},
which contributes to the neutrino mass at the tree level. The cosmological bound on neutrino masses $\sum_i m_{\nu_i}\lesssim 1\mathrm{eV}$~\cite{Ade:2013zuv}
leads to the bound on this fermionic bilinear R-parity violating effect~\cite{Barbier:2004ez}
\begin{equation}
\sin\xi \lesssim 3\times 10^{-6}\sqrt{1+\tan^2\beta} \left(\frac{M_2}{100\mathrm{GeV}}\right)^{1/2}, \label{bilinear_sinxi}
\end{equation}
where $M_2$ is the mass of $SU(2)_L$ gaugino, and
$\tan\beta=v_u/v_d$ is the ratio between vacuum expectation values (VEVs) of two Higgs fields $v_u$ and $v_d$.
In the basis used in the review [Ref.~\cite{Barbier:2004ez}] where VEVs of sneutrino fields vanish and Yukawa couplings of charged leptons become diagonal,
we have the relation $\sin^2\xi = \sum_i\mu_i^2/\mu^2$, where $\mu$ is the coefficient of the ``$\mu$-term" in the MSSM superpotential
(i.e. $W_{\rm MSSM}\subset \mu H_uH_d$).
With the choice of this specific basis, we can rewrite the constraint [Eq.~\eqref{bilinear_sinxi}] as
\begin{equation}
\sum_i\mu_i^2 \lesssim 9\times 10^{-12}(1+\tan^2\beta)\left(\frac{M_2}{100\mathrm{GeV}}\right)\mu^2. \label{bilinear_mui}
\end{equation}

In a similar way, the effect of the bilinear R-parity violation in the sfermion sector can be parametrized by the basis-independent quantity $\sin\zeta$,
which contributes to the neutrino mass at the one-loop level~\cite{Davidson:2000ne}.
Again, the cosmological bound $\sum_i m_{\nu_i}\lesssim 1\mathrm{eV}$ leads to the constraint~\cite{Barbier:2004ez},
$\sin\zeta \lesssim (10^{-4}-10^{-3})(m_{\tilde{f}}/100\mathrm{GeV})^{3/2}(100\mathrm{GeV}/\sqrt{B})^2$,
where $B$ is the coefficient of the ``B-term" $\mathcal{L}_{\rm soft}\subset -(BH_u H_d+\mathrm{h.c.})$
in the MSSM Lagrangian.

%%%%%%%%%%%%%%%%%%%%%%%%%%%%%%%%%%%%%%%%%%%%%%%%%%%%%%%%%%%%%%%%
\subsection{\label{sec2-5}Other indirect bounds}
%%%%%%%%%%%%%%%%%%%%%%%%%%%%%%%%%%%%%%%%%%%%%%%%%%%%%%%%%%%%%%%%
In addition to the bounds described in the previous subsections, the magnitude of R-parity violating couplings can be constrained
from various experimental results, such as charged and neutral current interactions, CP violations, electroweak precision measurements,
hadron or lepton flavor violating processes, and B or L violating processes.
These experimental results generically lead to the mild bounds as~\cite{Barbier:2004ez}
\begin{equation}
\lambda <\mathcal{O}(10^{-3}-10^{-1})\left(\frac{m_{\tilde{f}}}{100\mathrm{GeV}}\right).
\end{equation}
An exception is the observation of $n-\bar{n}$ oscillation, which gives a rather tight bound on the specific couplings~\cite{Zwirner:1984is,Barbier:2004ez}
\begin{equation}
|\lambda_{11k}''| \lesssim \mathcal{O}(10^{-8}-10^{-7})\frac{10^8\mathrm{sec}}{\tau_{\rm osc}}\left(\frac{m_{\tilde{f}}}{100\mathrm{GeV}}\right)^{5/2},
\end{equation}
where $\tau_{\rm osc}$ is the oscillation time.
Another bound on the baryonic R-parity violation can be obtained from the observation of $NN\to KK$ transition~\cite{Goity:1994dq},
which leads to $|\lambda_{121}''|\lesssim\mathcal{O}(10^{-7}-10^0)$,
but this result contains large uncertainties because of the dependence on hadronic and nuclear structure inputs.

If the magnitude of R-parity violating couplings are extremely small, the lifetime of the LSP exceeds the present age of the universe.
The decay of such long-lived LSPs via R-parity violating couplings is severely constrained, since it contributes to the observed cosmic ray flux.
In Ref.~\cite{Baltz:1997ar}, the following bounds are obtained from the observed antiproton flux
\begin{equation}
|\lambda_{ijk}'|,\ |\lambda_{ijk}''| \lesssim \mathcal{O}(10^{-18}). \label{long-lived_LSP_antiproton}
\end{equation}
The observed positron flux also puts the following bounds on the R-parity violating couplings~\cite{Berezinsky:1996pb}
\begin{align}
|\lambda_{ijk}|,\ |\lambda_{ijk}'|,\ |\lambda_{ijk}''| &< 4\times 10^{-21}Z_{\chi\tilde{H}}^{-1}\left(\frac{m_{\tilde{f}}}{1\mathrm{TeV}}\right)^2\left(\frac{100\mathrm{GeV}}{m_{\rm LSP}}\right)^{9/8}\left(\frac{1\mathrm{GeV}}{m_f}\right)^{1/2},\nonumber\\
\mu_i &< 6\times 10^{-23}\mathrm{GeV}Z_{\chi\tilde{H}}^{-1}\left(\frac{100\mathrm{GeV}}{m_{\rm LSP}}\right)^{7/4},  \label{long-lived_LSP_positron}
\end{align}
where $m_f$ is the mass of the fermion emitted from the decay process, and $Z_{\chi\tilde{H}}$ is the amount of the Higgsino components in the neutralino LSP.
Since the parameter region satisfying Eqs.~\eqref{long-lived_LSP_antiproton} and~\eqref{long-lived_LSP_positron} is identical to the R-parity conserving model from the cosmological point of view,
we will not consider such a case in this paper.

%%%%%%%%%%%%%%%%%%%%%%%%%%%%%%%%%%%%%%%%%%%%%%%%%%%%%%%%%%%%%%%%
\section{\label{sec3}Affleck-Dine mechanism}
%%%%%%%%%%%%%%%%%%%%%%%%%%%%%%%%%%%%%%%%%%%%%%%%%%%%%%%%%%%%%%%%
In this section, we turn our attention to the mechanism for baryogenesis.
In the AD mechanism, A-terms associated with B or L number violating operators in the superpotential play a crucial role to generate the B asymmetry.
The usual way considered in the literature is to assume some nonrenormalizable operator in the superpotential as an origin of these A-terms.
In the following, we investigate whether the AD mechanism works if we admit the existence of \emph{renormalizable} R-parity violating operators [Eq.~\eqref{RPV_superpotential}]
instead of the nonrenormalizable terms.
Note that every term in R-parity violating superpotential shown in Eq.~\eqref{RPV_superpotential} ($H_uL_i$, $L_iL_jE^c_k$, $L_iQ_jD_k^c$, and $U_i^cD_j^cD_k^c$)
corresponds to the flat direction of the MSSM in the absence of the R-parity violation~\cite{Gherghetta:1995dv}.
Therefore, a small R-parity violating coupling lifts these flat directions in the MSSM, which induces the AD mechanism.
In the remaining part of this section,
first we derive the potential for the flat direction in Sec.~\ref{sec3-1}.
Based on this setup, in Sec.~\ref{sec3-2} we solve the evolution of the scalar field condensation after inflation, 
and estimate the amount of the B asymmetry generated in this scenario.

%%%%%%%%%%%%%%%%%%%%%%%%%%%%%%%%%%%%%%%%%%%%%%%%%%%%%%%%%%%%%%%%
\subsection{\label{sec3-1}Potential for the Affleck-Dine field}
%%%%%%%%%%%%%%%%%%%%%%%%%%%%%%%%%%%%%%%%%%%%%%%%%%%%%%%%%%%%%%%%
The form of the potential for a flat direction relevant to the AD mechanism, which we call the AD field hereafter, 
might depend on the scenario of the supersymmetry breaking.
In this paper, we work in the framework of gravity (or anomaly) mediated supersymmetry breaking in the context of F-term inflation~\cite{Lyth:1998xn}.
If the mass of the gravitino is sufficiently small, we must take account of the effect of gauge mediated supersymmetry breaking,
in which the scalar potential is modified and the estimation of the B asymmetry becomes nontrivial~\cite{deGouvea:1997tn}.
However, as we will see in Sec.~\ref{sec5}, the present B asymmetry can be explained for the case where gravitino is heavier than $m_{3/2} \gtrsim \mathcal{O}(10^4)\mathrm{GeV}$.
Therefore, in the following we do not consider the scenario with gauge mediated supersymmetry breaking.

Let us consider the following superpotential:
\begin{equation}
W = W_{\rm inf}(I) + W_{\not{R_p}}(\phi),
\end{equation}
where $W_{\rm inf}(I)$ is the superpotential for the inflaton field $I$, and $W_{\not{R_p}}(\phi)$ is the R-party violating interaction from Eq.~\eqref{RPV_superpotential}
for a flat direction parametrized by the field $\phi$.
For example, when we consider the combination $L_1L_2E^c_2$, we parametrize
\begin{equation}
L_1 = \frac{1}{\sqrt{3}}\left(
\begin{array}{c}
\phi \\
0
\end{array}
\right), \qquad
L_2 = \frac{1}{\sqrt{3}}\left(
\begin{array}{c}
0 \\
\phi 
\end{array}
\right), \qquad
E_2^c = \frac{1}{\sqrt{3}}\phi,
\end{equation}
where the columns represent $SU(2)_L$ components.
This direction satisfies both F-flat and D-flat conditions in the absence of R-parity violation~\cite{Gherghetta:1995dv},
but it is lifted by the existence of the R-parity violating interaction, which we denote as $W_{\not{R_p}}(\phi) = \lambda_{122}\phi^3/3$.
Similar arguments can be applied for other trilinear combinations $L_iQ_jD_k^c$ and $U_i^cD_j^cD_k^c$, and hence
we use the following superpotential to represent the R-parity violating effects:
\begin{equation}
W_{\not{R_p}}(\phi) = \frac{1}{3}\lambda\phi^3, \label{W_RPV_trilinear}
\end{equation}
where $\lambda$ stands for any of the trilinear R-parity violating couplings ($\lambda_{ijk}$, $\lambda_{ijk}'$, or $\lambda_{ijk}''$).

We note that the effect of the bilinear R-parity violating terms $\mu_iH_uL_i$ is negligible.
As mentioned in Sec.~\ref{sec2-4}, the size of the bilinear R-parity violating effects is tightly constrained from the cosmological bound
on neutrino masses. This roughly corresponds to the bound $\mu_i\lesssim \mathcal{O}(10^{-6})\mu$ [see Eq.~\eqref{bilinear_mui}].
This fact implies that the magnitude of $\mu_i$-term, which supplies CP violation for the AD mechanism,
is smaller than that of the scalar mass term by a factor of $\mathcal{O}(10^{-6})$.
This situation should be compared with the minimal scenario using the $H_uL_i$ direction~\cite{Moroi:1999uc},
where the AD leptogenesis occurs when the magnitude of the soft mass term becomes comparable with that of the A-term.
In other words, we expect that the efficiency of the leptogenesis with the bilinear R-parity violating terms
is smaller than that of the scenario in Ref.~\cite{Moroi:1999uc} by a factor of $\mathcal{O}(10^{-6})$.
Since it is difficult to explain the baryon asymmetry observed today in such a scenario, we do not consider the bilinear R-parity violating
terms and concentrate on the trilinear terms represented by Eq.~\eqref{W_RPV_trilinear}.

The effective scalar potential in supergravity is given by~\cite{Nilles:1983ge}
\begin{align}
V = e^{K/M_{\rm Pl}^2}\left[(D_aW)(K^{-1})^a_b(D^bW^*) - \frac{3}{M_{\rm Pl}^2}|W|^2\right],
\end{align}
with
\begin{align}
D_aW = \frac{\partial W}{\partial \phi^a} + \frac{W}{M_{\rm Pl}^2}\frac{\partial K}{\partial \phi^a}, \qquad 
D^{a}W^* = \left(\frac{\partial W}{\partial \phi^a}\right)^* + \frac{W^*}{M_{\rm Pl}^2}\frac{\partial K}{\partial \phi^{\dagger}_a},\qquad
K^b_a = \frac{\partial^2 K}{\partial \phi^{\dagger}_b\partial\phi^a},
\end{align}
where the indices $a,b=I,\phi$ stand for the fields involved in the model (i.e. the inflaton field $I$ and the AD field $\phi$),
$K=K(\phi^a,\phi^{\dagger}_a)$ is the K\"{a}hler potential which will be specified later, and $M_{\rm Pl}=(8\pi G)^{-1/2}=2.4\times 10^{18}\mathrm{GeV}$ is the Planck mass.
Let us consider the contribution of $W_{\rm inf}$ to the scalar potential,
\begin{align}
V \simeq e^{K/M_{\rm Pl}^2}\left(F_I^*F^I - \frac{3}{M_{\rm Pl}^2}|W_{\rm inf}(I)|^2\right),
\end{align}
where $F_I^*F^I = D_IW_{\rm inf}(K^{-1})^I_I D^IW_{\rm inf}^*$. Suppose that the K\"{a}hler potential takes the following form:
\begin{equation}
K = \phi^{\dagger}\phi + I^{\dagger}I + \frac{a}{M_{\rm Pl}^2}\phi^{\dagger}\phi I^{\dagger}I + \dots,
\end{equation}
where $a$ is some numerical coefficient.
The scalar potential reduces to
\begin{align}
V& \sim e^{K/M_{\rm Pl}^2}\left[|F_I|^2\frac{1+\frac{a}{M_{\rm Pl}^2}|I|^2}{1+\frac{a}{M_{\rm Pl}^2}(|\phi|^2+|I|^2)}-\frac{3}{M_{\rm Pl}^2}|W_{\rm inf}(I)|^2\right] \nonumber\\
&\supset |\phi|^2\left[\frac{|F_I|^2}{M_{\rm Pl}^2}\left\{(1-a)+(1+a^2)\frac{|I|^2}{M_{\rm Pl}^2}\right\} - \frac{3}{M_{\rm Pl}^4}\left\{1+(1+a)\frac{|I|^2}{M_{\rm Pl}^2}\right\}|W_{\rm inf}(I)|^2\right]. \label{V_from_inf}
\end{align}
The magnitude of the second term $3|W_{\rm inf}|^2/M_{\rm Pl}^4$ in the square brackets
in Eq.~\eqref{V_from_inf} should be smaller than that of the first term $|F_I|^2/M_{\rm Pl}^2 \sim H^2$. Therefore, we have the negative Hubble mass squared for $\phi$ if $a \gtrsim 1$ is satisfied. Adding the usual soft supersymmetry breaking contribution $m^2_{\phi}|\phi|^2$, we describe the mass term for the AD field as
\begin{equation}
V_{\rm soft\ mass}(\phi) \simeq (m_{\phi}^2-cH^2)|\phi|^2,
\end{equation}
where $c\ge 0$ is a numerical coefficient.\footnote{A situation where there is a large Hubble-induced mass ($c\gg 1$)
and the soft mass is much larger than the coefficient of the A-term is considered in Ref.~\cite{Higaki:2012ba}.}

Next, let us consider the contributions from the terms containing $W_{\not{R_p}}$.
These include the interferences between $W_{\not{R_p}}$ and $W_{\rm inf}$ derived from the terms such as
$(K^{-1})^{\phi}_{\phi}(D_{\phi}W)(D^{\phi}W^*)$, $(K^{-1})^I_I(D_I W)(D^IW^*)$, $-3|W|^2/M_{\rm Pl}^2$,
and $(K^{-1})^{\phi}_I(D_{\phi}W)(D^IW^*)+\mathrm{h.c.}$
For the case where the inflaton takes a large value $|I|\sim M_{\rm Pl}$, these terms can be written as
\begin{equation}
V(\phi) \sim HM_{\rm Pl}^3 g(\phi/M_{\rm Pl}) + \mathrm{h.c.},
\end{equation}
where $g$ is some polynomial of $\phi/M_{\rm Pl}$. In general, this contribution is suppressed if the inflaton takes $|I|\ll M_{\rm Pl}$.
However, if there exist the following terms in the K\"ahler and superpotential:
\begin{align}
K' &= a'\frac{I}{M_{\rm Pl}}\phi^{\dagger}\phi + \mathrm{h.c.}, \nonumber \\
W' &= b'\frac{I}{M_{\rm Pl}}W_{\not{R_p}},
\label{K_prime_W_prime}
\end{align}
with some numerical coefficients $a'$ and $b'$,
we see that the scalar potential contains the following terms:
\begin{equation}
\left(a'\frac{\partial W_{\not{R_p}}}{\partial \phi}\phi + b'W_{\not{R_p}}\right)\frac{F_I^*}{M_{\rm Pl}} + \mathrm{h.c.},
\label{Hubble_Aterms}
\end{equation}
which are not suppressed even if $|I|\ll M_{\rm Pl}$.
Together with a supersymmetry breaking effect $\sim m_{3/2}W_{\not{R_p}}(\phi)$ with $m_{3/2}$ being the gravitino mass,
we can write these contributions as the following form:
\begin{equation}
V_{\rm A\mathchar`-terms}(\phi) = a_h HW_{\not{R_p}}(\phi) + a_m m_{3/2}W_{\not{R_p}}(\phi) + \mathrm{h.c.},
\end{equation}
where $a_h$ and $a_m$ are numerical coefficients of $\mathcal{O}(1)$.

Finally, F-term contribution of the scalar $\phi$ leads to the following quartic interaction:
\begin{equation}
V_{F_{\phi}}(\phi) = \lambda^2|\phi|^{4}.
\end{equation}
Combining all the above ingredients, we obtain the following form of the potential for the AD field $\phi$:
\begin{align}
V(\phi) = (m_{\phi}^2 - cH^2)|\phi|^2 + \left(\frac{\lambda}{3} a_m m_{3/2}\phi^3 + \mathrm{h.c.}\right)
+ \left(\frac{\lambda}{3} a_h H\phi^3 + \mathrm{h.c.}\right)
+ \lambda^2|\phi|^{4}.
\end{align}
Because of the existence of the Hubble-induced A-term $a_hHW_{\not{R_p}}$,
the phase direction of $\phi$ acquires a mass of $\mathcal{O}(H)$ during inflation.
This suppresses the unwanted baryonic isocurvature fluctuations~\cite{Yokoyama:1993gb,Enqvist:1998pf,Enqvist:1999hv,Kawasaki:2001in,Kasuya:2008xp}
which are tightly constrained from current observations~\cite{Ade:2013uln}.\footnote{
If the operators shown in Eqs.~\eqref{K_prime_W_prime}
are not suppressed, the inflaton field $I$ must be a singlet under the R-symmetry.
This fact might lead to a problem in constructing the inflationary model, since in most models the R-symmetry remains to be a good symmetry during inflation~\cite{Kasuya:2008xp}.}
 
We note that the effect of the term proportional to $a_h$
can be omitted after inflation~\cite{Kasuya:2008xp}.
After the end of inflation, the inflaton field $I$ begins to oscillate around the minimum $I_{\rm min}$
so that the K\"{a}hler and superpotential can be written as
\begin{align}
K &= |I|^2 + \dots = I_{\rm min}^*\hat{I} + I_{\min}\hat{I}^{\dagger} + |\hat{I}|^2+ \dots, \nonumber\\
W &= \frac{1}{2}M_{\rm inf}(I-I_{\rm min})^2 + \dots = \frac{1}{2}M_{\rm inf}\hat{I}^2+ \dots,
\end{align}
where $\hat{I} = I - I_{\rm min}$ and $M_{\rm inf}$ is the mass of the inflaton.
Since the term with $a_h$ is
proportional to the rapidly oscillating factor $F_I = -M_{\rm inf}\hat{I}^*$ [see Eq.~\eqref{Hubble_Aterms}], 
this term can be dropped as long as the oscillation period $M_{\rm inf}^{-1}$
is much shorter than the Hubble time $H^{-1}$.
Omitting the $a_h$-term, we have 
\begin{equation}
V(\phi) = (m_{\phi}^2 - cH^2)|\phi|^2 + \left(\frac{\lambda}{3} a_m m_{3/2}\phi^3 + \mathrm{h.c.}\right) + \lambda^2|\phi|^4,
\label{scalar_potential}
\end{equation}
which holds after inflation.

%%%%%%%%%%%%%%%%%%%%%%%%%%%%%%%%%%%%%%%%%%%%%%%%%%%%%%%%%%%%%%%%
\subsection{\label{sec3-2}Dynamics of the Affleck-Dine field}
%%%%%%%%%%%%%%%%%%%%%%%%%%%%%%%%%%%%%%%%%%%%%%%%%%%%%%%%%%%%%%%%
During inflation, $\phi$ is frozen at the value determined by the minimum of the potential~\eqref{scalar_potential}.
Ignoring the soft mass term and the A-term, we find the minimum of the potential as
\begin{equation}
\phi_{\rm min} = \frac{\sqrt{c}H}{\sqrt{2}\lambda}. \label{phi_min}
\end{equation}
Let us consider the evolution of the homogeneous field value $\phi(t)$ after inflation.
Just after inflation, the soft mass term and the A-term are still negligible compared with the negative Hubble mass term,
and the equation of motion for the field $\phi$ is given by
\begin{equation}
\ddot{\phi} + 3H\dot{\phi} + \phi(-cH^2 + 2\lambda^2|\phi|^2) = 0, \label{EOM_before_osc}
\end{equation}
where a dot represents a derivative with respect to the cosmic time $t$.
Assuming the matter-dominated background $H=2/3t$, and denoting $\phi(t) = \chi(t)\phi_{\rm min}(t)$, 
we rewrite Eq.~\eqref{EOM_before_osc} as
\begin{equation}
\ddot{\chi} + \frac{4c}{9t^2}\chi(\chi^2-1) = 0.
\end{equation}
This equation implies that the evolution of $\chi$ can be understood as the classical motion in the potential
$V(\chi)= \frac{c}{9t^2}(\chi^4-2\chi^2)$.
Since the initial value is $\chi=1$ (e.g. $\phi=\phi_{\rm min}$) and $V(\chi)$ has a minimum at the same location,
we see that $\chi$ always stay in the value $\chi=1$.
Therefore $\phi$ tracks the value $\phi_{\rm min}$ given by Eq.~\eqref{phi_min} after inflation.\footnote{
The recent detection of the tensor mode reported in Ref.~\cite{Ade:2014xna}
indicates that the Hubble parameter during inflation becomes $H_{\rm inf} =\mathcal{O}(10^{14})\mathrm{GeV}$.
In this case the value for $\phi_{\rm min}$ given by Eq.~\eqref{phi_min} exceeds the Planck scale for $\lambda < \mathcal{O}(10^{-4})$ during inflation.
Therefore, it is presumed that the AD field takes a value of $\mathcal{O}(M_{\rm Pl})$ at the end of inflation.
Afterwords, it oscillates around the minimum with the averaged value tracking the value shown in Eq.~\eqref{phi_min}.
This fact does not affect the estimation for the baryon asymmetry, which is determined by the field value at the onset of the oscillation $\phi(t_{\rm osc})\propto H_{\rm osc}/\lambda$
[see Eq.~\eqref{n_t_osc}].
Furthermore, if the value of the AD field is as large as $|\phi_{\rm inf}|\simeq M_{\rm Pl}$ during inflation,
the magnitude of the fluctuations in the phase direction of $\phi$ becomes
$\Delta_{\delta\theta}=H_{\rm inf}/(2\pi|\phi_{\rm inf}|) \simeq \mathcal{O}(10^{-5})$, which marginally avoids the constraint
from baryonic isocurvature fluctuations~\cite{Kawasaki:2008jy}.}
Eventually it begins to oscillate when the soft term becomes relevant ($H\sim m_{\phi}$).

The B/L number stored in the AD field can be quantified by the following charge density:
\begin{equation}
n= i(\dot{\phi}^*\phi-\phi^*\dot{\phi}).
\end{equation}
According to the choice of the flat direction, this quantity is related to the B number $n_B$ or L number $n_L$ as follows:
\begin{align}
n_B = 0, \quad  n_L = \frac{1}{3}n, \quad &\mathrm{for}\quad L_iL_jE_k^c \quad \mathrm{or} \quad L_iQ_jD_k^c, \nonumber\\
n_B = -\frac{1}{3}n, \quad n_L = 0, \quad &\mathrm{for}\quad U_i^cD_j^cD_k^c. \label{relation_between_n-nB-nL}
\end{align}
Let us compute $n$ explicitly.
Using the equation of motion for $\phi$, we have
\begin{equation}
\frac{d}{dt}(nR^3) = 2 R^3\mathrm{Im}\left(\frac{\partial V}{\partial\phi}\phi\right),
\end{equation}
where $R$ is the scale factor of the universe.
Performing the time integration from the end of inflation $t_{\rm inf}$ with the potential~\eqref{scalar_potential}, we obtain
\begin{align}
R(t)^3n(t) &= 2 \int^t_{t_{\rm inf}}dt R^3\mathrm{Im}\left(\lambda a_m m_{3/2} \phi^3\right) \nonumber\\
&= 2 \int^{t_{\rm osc}}_{t_{\rm inf}}dt R^3\left(\lambda |a_m| m_{3/2} |\phi|^3\delta_{\rm eff}\right)
+ 2 \int^t_{t_{\rm osc}}dt R^3\left(\lambda |a_m| m_{3/2} |\phi|^3\delta_{\rm eff}\right), \label{R^3n}
\end{align}
where $t_{\rm osc}$ is the time at the beginning of the oscillation [$H(t_{\rm osc})\simeq m_{\phi}$], and $\delta_{\rm eff}\equiv \sin(\arg (a_m)+3\arg(\phi))$.
The integrand scales as $R^3|\phi|^3\propto t^{-1}$ both for $t<t_{\rm osc}$ and $t>t_{\rm osc}$, which implies that the contribution from
the second term in Eq.~\eqref{R^3n} can be comparable with that from the first term. However, we expect that the contribution from the second term 
is insignificant since the sign of the phase factor $\delta_{\rm eff}$ changes rapidly after $\phi$ begins to oscillate.
From the first term of Eq.~\eqref{R^3n}, the charge density at the time $t=t_{\rm osc}$ is estimated as
\begin{align}
n(t_{\rm osc}) &\simeq 2\lambda|a_m|m_{3/2}\delta_{\rm eff}\frac{2}{3H_{\rm osc}}|\phi(t_{\rm osc})|^3\ln\frac{t_{\rm osc}}{t_{\rm inf}} \nonumber\\
& \simeq \frac{\sqrt{2 c^3}}{3\lambda^2}|a_m|m_{3/2}\delta_{\rm eff}H_{\rm osc}^2\ln\frac{t_{\rm osc}}{t_{\rm inf}}, \label{n_t_osc}
\end{align}
where $H_{\rm osc}$ is the Hubble parameter at the time $t=t_{\rm osc}$,
and we used the relation
$|\phi(t_{\rm osc})|=\sqrt{c}H_{\rm osc}/\sqrt{2}\lambda$ in the second line.
Deriving Eq.~\eqref{n_t_osc}, we neglected the time dependence of $\delta_{\rm eff}$, 
and treated it as a constant.

\begin{figure*}[htp]
\centering
$\begin{array}{cc}
\subfigure[]{
\includegraphics[width=0.45\textwidth]{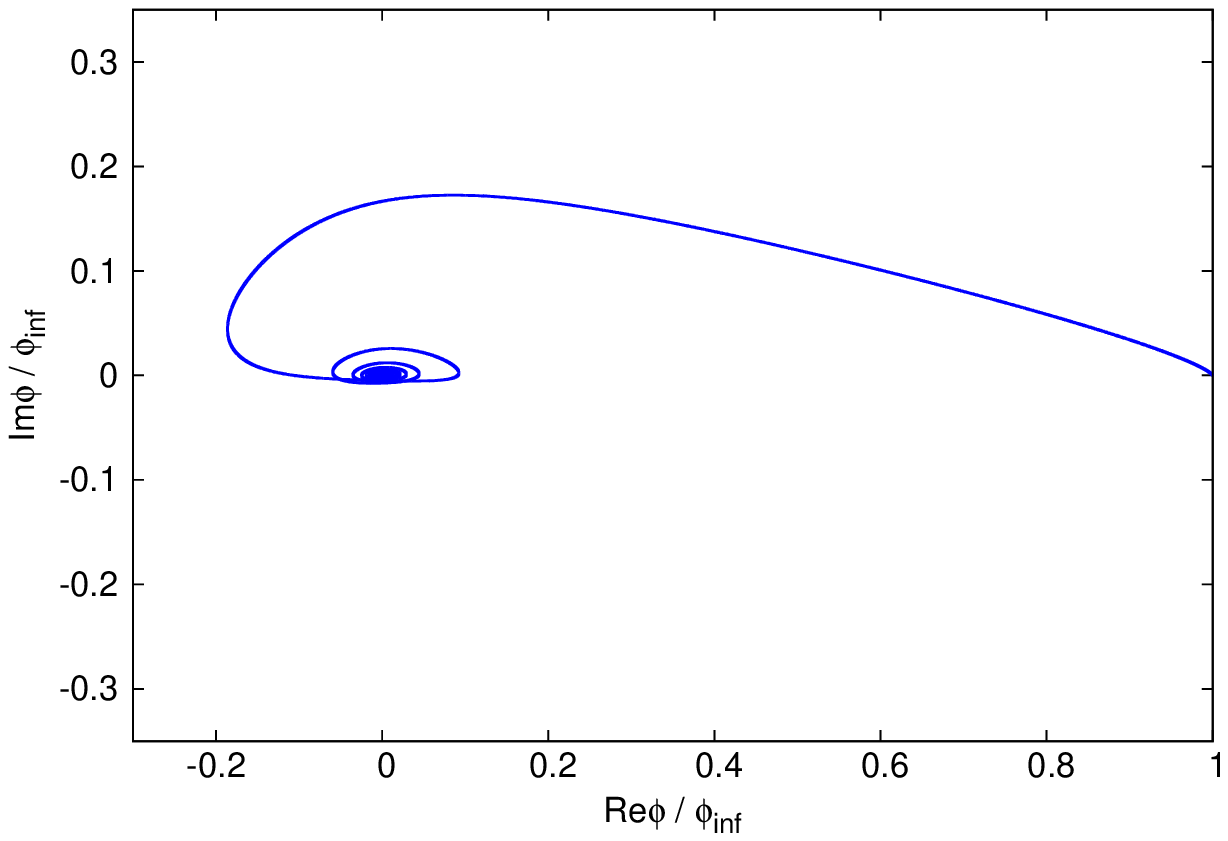}}
\hspace{20pt}
\subfigure[]{
\includegraphics[width=0.45\textwidth]{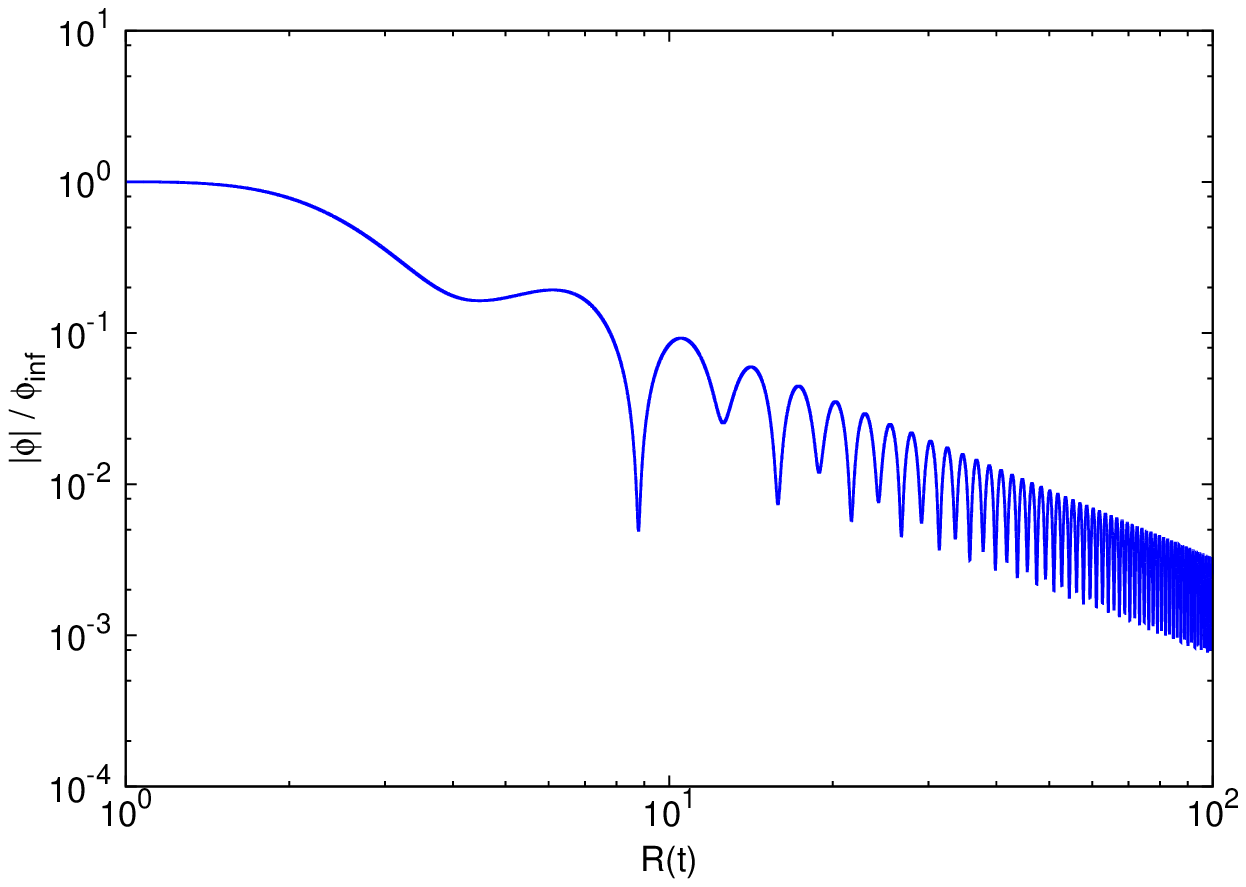}}
\end{array}$
\caption{Time evolution of the AD field after inflation.
Left panel (a) shows the trajectory of $\phi$ in the field space.
Right panel (b) shows the evolution of the amplitude $|\phi|$, where the horizontal axis corresponds to the scale factor $R(t)$.
In these figures, the field amplitudes are plotted in the unit of $\phi_{\rm inf} = \sqrt{c}H_{\rm inf}/\sqrt{2}\lambda$.
In numerical calculations we fixed the parameters as $\lambda =10^{-4}$, $c=1$, $H_{\rm inf}/m_{\phi}=4$, $|a_m|=1$,
and $\mathrm{arg}(a_m)=\pi/2$.}
\label{fig1}
\end{figure*}

To check the expression~\eqref{n_t_osc}, we integrate the right-hand side of Eq.~\eqref{R^3n} numerically by solving the classical field equation for the scalar field $\phi$.
Here, we solve the equation of motion for $\phi$ with the potential given by Eq.~\eqref{scalar_potential} in the matter-dominated background.
The initial condition is fixed as $\mathrm{Re}\phi = \phi_{\rm inf} \equiv \sqrt{c}H_{\rm inf}/\sqrt{2}\lambda$ and $\mathrm{Im}\phi=0$,
where $H_{\rm inf}$ is the Hubble parameter at the end of inflation.
Figure~\ref{fig1}~(a) shows the trajectory of the AD field in the complex plane $(\mathrm{Re}\phi,\mathrm{Im}\phi)$.
We see that the scalar field rotates around the phase direction, reducing its amplitude.
In Fig.~\ref{fig1}~(b), we also plot the evolution of the field amplitude $|\phi(t)|$, from which we confirmed the behavior $|\phi|\propto R^{-1}t^{-1/3}\propto R^{-3/2}$.

%%%%%%%%%%%%%%%%%%%%%%%%%%%fig1%%%%%%%%%%%%%%%%%%%%%%%%%%%%%%%%%%%
%%%%%%%%%%%%%%%%%%%%%%%%%%%%%%%%%%%%%%%%%%%%%%%%%%%%%%%%%%%%%%%%

\begin{figure}[htbp]
\begin{center}
\includegraphics[]{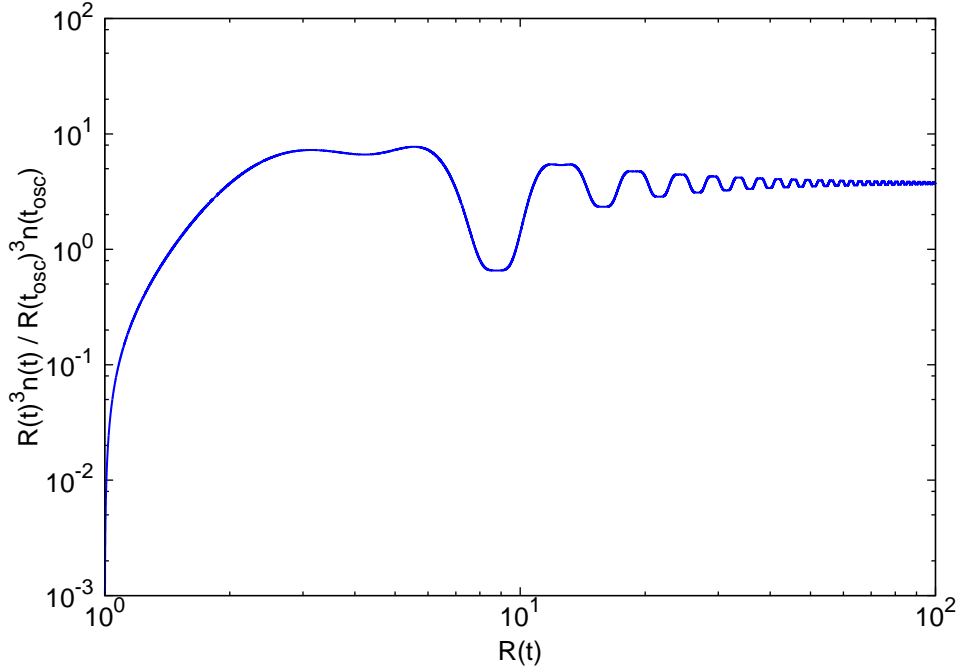}
\end{center}
\caption{Time evolution of the ratio between $R(t)^3n(t)$ [in Eq.~\eqref{R^3n}] 
and $R(t_{\rm osc})^3n(t_{\rm osc})$.
Here we used Eq.~\eqref{n_t_osc} with $\delta_{\rm eff}=1$ to represent $n(t_{\rm osc})$.
The horizontal axis corresponds to the scale factor $R(t)$.
In this plot, we used the same numerical data as Fig.~\ref{fig1}.}
\label{fig2}
\end{figure}

The time evolution of the integral in Eq.~\eqref{R^3n} is shown in Fig.~\ref{fig2}.
We confirmed that the integrated value approaches to the analytic estimation in Eq.~\eqref{n_t_osc} (with $\delta_{\rm eff}=1$)
after the beginning of the oscillation.
However, the convergence value gets a bit dislocated from what we expected in Eq.~\eqref{n_t_osc} by a factor of $\mathcal{O}(1)$.
It is probable that this discrepancy is caused by the uncertainty of the definition of $t_{\rm osc}$, at which we truncate the integral to obtain the
analytic expression [Eq.~\eqref{n_t_osc}].
Furthermore, the convergence value varies with $\mathrm{arg}(a_m)$ and $H_{\rm inf}/m_{\phi}$, which also gives an uncertainty of $\mathcal{O}(1)$.
On the other hand, we checked that the final value for the ratio $R(t)^3n(t)/R(t_{\rm osc})^3n(t_{\rm osc})$ hardly depends on other parameters
such as $\lambda$ and $|a_m|$.
Regarding these facts, hereafter we use Eq.~\eqref{n_t_osc} to estimate the amount of the B asymmetry
with the factor $\delta_{\rm eff}$ replaced by $\tilde{\delta}_{\rm eff}$, which contains some uncertainties such as dependences on $t_{\rm osc}$, $\mathrm{arg}(a_m)$, and $H_{\rm inf}/m_{\phi}$.

%%%%%%%%%%%%%%%%%%%%%%%%%%%fig2%%%%%%%%%%%%%%%%%%%%%%%%%%%%%%%%%%%
%%%%%%%%%%%%%%%%%%%%%%%%%%%%%%%%%%%%%%%%%%%%%%%%%%%%%%%%%%%%%%%%

The ratio of the charge density $n$ to the entropy density $s$ after reheating ($t=t_{\rm RH}$) is estimated as
\begin{align}
\frac{n}{s} &= \frac{1}{s(t_{\rm RH})}\left(\frac{R(t_{\rm osc})}{R(t_{\rm RH})}\right)^3 n(t_{\rm osc}) \nonumber\\
&\simeq \frac{\sqrt{2c^3}|a_m|\tilde{\delta}_{\rm eff}}{12\lambda^2}\frac{m_{3/2}T_{\rm RH}}{M_{\rm Pl}^2}, \label{n_ov_s}
\end{align}
where we used Eq.~\eqref{n_t_osc} with $\ln(t_{\rm osc}/t_{\rm inf})\simeq 1$ in the second equality, and $T_{\rm RH}$ is the reheating temperature.
We see that the ratio $n/s$ does not depend on the Hubble parameter at the beginning of the oscillation $H_{\rm osc}$.
In the case where primordial L asymmetry is generated via $L_iL_jE_k^c$ or $L_iQ_jD_k^c$ direction,
one must take account of the conversion effect between L and B numbers due to the sphaleron interactions~\cite{Kuzmin:1985mm}.
Including this effect, we can estimate the present B asymmetry in terms of the primordial L asymmetry~\cite{Khlebnikov:1988sr,Harvey:1990qw}:
\begin{equation}
\frac{n_B}{s} = -\frac{8}{23}\frac{n_L}{s}.
\end{equation}
Together with the factors shown in Eqs.~\eqref{relation_between_n-nB-nL}, we obtain the following result for the present B asymmetry,
\begin{equation}
\frac{n_B}{s} = \gamma \frac{\sqrt{2c^3}|a_m|\tilde{\delta}_{\rm eff}}{12\lambda^2}\frac{m_{3/2}T_{\rm RH}}{M_{\rm Pl}^2},
\end{equation}
where the coefficient $\gamma$ is given by\footnote{At this stage, we do not care about the overall sign of $n_B/s$, which
should be determined by the sign of CP violation.}
\begin{equation}
\gamma = 
\left\{
\begin{array}{ll}
8/69 & \mathrm{for}\quad L_iL_jE_k^c \quad \mathrm{or} \quad L_iQ_jD_k^c, \\
1/3 & \mathrm{for}\quad U_i^cD_j^cD_k^c.
\end{array}
\right.
\end{equation}
If we take the values of numerical coefficients as $c=|a_m|=\tilde{\delta}_{\rm eff}=1$, we have
\begin{equation}
\frac{n_B}{s} \simeq 2 \times 10^{-9}\gamma\left(\frac{10^{-10}}{\lambda}\right)^2\left(\frac{T_{\rm RH}}{10^5\mathrm{GeV}}\right)
\left(\frac{m_{3/2}}{10\mathrm{TeV}}\right).
\label{n_to_s_final}
\end{equation}
Note that a large baryon asymmetry is obtained if $\lambda$ is \emph{small}.
For a smaller value of $\lambda$, the AD field acquires a large expectation value [Eq.~\eqref{phi_min}], which enhances the amount
of the primordial B/L asymmetry.

Before closing this section, let us comment on the magnitude of the A-term.
If the magnitude of $|a_m|$ is sufficiently large, 
the radial direction $|\phi|$ would be trapped in the local minimum of the potential before it begins to oscillate,
which spoils the AD mechanism.
Let us consider the potential for the radial direction $|\phi|$, on which the coefficient of the A-term becomes negative
[i.e. $\cos(\arg(a_m)+3\arg(\phi))=-1$]:
\begin{equation}
V(|\phi|) = m_{\phi}^2|\phi|^2 - \frac{2}{3}\lambda|a_m| m_{3/2} |\phi|^3 + \lambda^2|\phi|^4,
\end{equation}
where we ignored the Hubble mass term.
The condition for the absence of the local minimum is given by
\begin{equation}
|a_m| m_{3/2} < 2\sqrt{2}m_{\phi}.
\end{equation}
It is satisfied for $|a_m|\lesssim 2\sqrt{2}$, if the gravitino mass and the soft mass are comparable.
Note that the term of the form $\sim m_{3/2}W_{\not{R_p}}(\phi)$ exactly vanishes in the framework of supergravity
if $W_{\not{R_p}}(\phi)$ corresponds to the polynomial of third order in $\phi$.
In order to guarantee the existence of such a contribution, it is necessary to assume some couplings of the form like $W \sim (Z/M_{\rm Pl})L_iL_jE_k^c$
with a moduli field $Z$ in the bare superpotential~\cite{Kaplunovsky:1993rd}.
This kind of moduli field typically acquires a large expectation value with a small mass, being abundant in the early universe,
which might spoil the standard cosmological scenario~\cite{Coughlan:1983ci,Ellis:1986zt,deCarlos:1993jw,Banks:1993en}.
Alternatively, we can assume that the A-term is generated via the effect of anomaly mediated supersymmetry breaking~\cite{Randall:1998uk,Giudice:1998xp}
to avoid this cosmological problem,
but in this case the magnitude of the A-term is loop suppressed such that $|a_m|\simeq \mathcal{O}(10^{-2})$.
In this paper, we just assume that these subtleties are resolved, and estimate the present B asymmetry by fixing $|a_m|=1$.

%%%%%%%%%%%%%%%%%%%%%%%%%%%%%%%%%%%%%%%%%%%%%%%%%%%%%%%%%%%%%%%%
\section{\label{sec4}Finite temperature effects and Q-balls}
%%%%%%%%%%%%%%%%%%%%%%%%%%%%%%%%%%%%%%%%%%%%%%%%%%%%%%%%%%%%%%%%
After the AD field begins to oscillate, the field condensate develops an instability~\cite{Kusenko:1997si}
to form nontopological solitons called Q-balls~\cite{Coleman:1985ki}.
Through this process, the charge (i.e. B or L number) stored in the AD field is converted in the form of Q-balls.
Therefore, it is important to investigate the cosmological evolution of Q-balls in order to estimate the present B asymmetry.
In particular, there is a possibility that long-lived Q-balls preserve the primordial B number
against the sphaleron erasure effect even in the presence of a large R-parity violation.
Here, we mainly consider the following two possibilities:
\begin{enumerate}
\item {\it Small $\lambda$ scenario}.
If the R-parity violating coupling is sufficiently small to satisfy Eq.~\eqref{sphaleron_lambda},
the sphaleron erasure effect
is ineffective, and we can evaluate the baryon number in the usual way.
In this case, there is a lower bound on $\lambda$ obtained from the requirement that
the decay of unstable LSPs does not spoil the standard BBN scenario [Eq.~\eqref{light_element_lambda}].
\item {\it Large $\lambda$ scenario}.
If the R-parity violating coupling is larger than the bound [Eq.~\eqref{sphaleron_lambda}],
B/L number violating interactions remain in thermal equilibrium until
the temperature becomes lower than $T_{\rm erasure}\approx\mathcal{O}(m_{\tilde{f}})$.
As was discussed in Sec.~\ref{sec2-3}, together with the sphaleron effect, they wash out the B/L number produced before
the EWPT.
However, it is possible to generate the baryon number due to the decay of Q-balls, if they have survived
until the epoch where equilibrium R-parity violating interactions have become irrelevant.
\end{enumerate}

In order to discuss the validity of the large $\lambda$ scenario, it is necessary to clarify whether Q-balls live until the 
time when the sphaleron erasure effect can be neglected.
Here we must take account of two effects: One is the evaporation of Q-balls into the surrounding plasma~\cite{Laine:1998rg,Banerjee:2000mb}.
Another is the effect caused by the existence of the $U(1)$ violating A-term, which makes Q-balls unstable~\cite{Kawasaki:2005xc}.
Because of these two effects, Q-balls are likely to be destroyed.
Furthermore, conditions for the survival of Q-balls become more complicated if we include the finite temperature corrections to the potential of the AD field.
In this section, we first introduce the finite temperature effects in Sec.~\ref{sec4-1}.
Then we estimate the charge of Q-balls and the condition for their evaporation in Sec.~\ref{sec4-2}.
Finally, the stability condition for Q-balls is discussed in Sec.~\ref{sec4-3}.

%%%%%%%%%%%%%%%%%%%%%%%%%%%%%%%%%%%%%%%%%%%%%%%%%%%%%%%%%%%%%%%%
\subsection{\label{sec4-1}Early oscillation due to the finite temperature effects}
%%%%%%%%%%%%%%%%%%%%%%%%%%%%%%%%%%%%%%%%%%%%%%%%%%%%%%%%%%%%%%%%
It is known that the finite temperature effects change the dynamics of the AD field, if the reheating temperature is sufficiently high~\cite{Allahverdi:2000zd}.
We must take account of 
two effects coming from one-loop order and two-loop order.
At the one-loop order, the existence of some fields $\psi_k$ which are in the thermal bath and directly couple to
the AD field induces the thermal mass of the form~\cite{Allahverdi:2000zd,Asaka:2000nb}
\begin{equation}
V(\phi) \supset c_k f_k^2 T^2|\phi|^2, \label{Vphi_thermal_mass}
\end{equation}
where $c_k=\mathcal{O}(0.1\mathchar`-1)$ is a constant determined by degrees of freedom of the field $\psi_k$,
and $f_k$ represents the magnitude of the (gauge or Yukawa) coupling between $\phi$ and $\psi_k$, which takes a value of $\mathcal{O}(10^{-5}-1)$ in the supersymmetric SM~\cite{Asaka:2000nb}.
Note that this effect is exponentially suppressed for $f_k|\phi|>T$.
On the other hand, if $|\phi|\gg T$, fields which do not directly couple with the AD field induce the following term through
the effective gauge coupling at the two-loop order~\cite{Anisimov:2000wx,Fujii:2001zr}
\begin{equation}
V(\phi) \supset a_g\alpha_g(T)^2T^4\ln\left(\frac{|\phi|^2}{T^2}\right), \label{Vphi_thermal_log}
\end{equation}
where $a_g$ is a numerical coefficient of $\mathcal{O}(1)$, $\alpha_g(T)=g(T)^2/4\pi$, and $g(T)$ is the gauge (or Yukawa) coupling evaluated at the scale $T$.
When we take account of these effects, the condition that the AD field starts to oscillate is given by
\begin{equation}
H^2 \simeq m_{\phi}^2 + \sum_{f_k|\phi|<T}c_k f_k^2T^2 + \frac{1}{|\phi|^2}a_g\alpha_g^2T^4. \label{start_osc}
\end{equation}

Let us consider the case where the thermal mass term [Eq.~\eqref{Vphi_thermal_mass}] dominates the right-hand side of Eq.~\eqref{start_osc}.
In this case, the oscillation of $\phi$ starts when both of the following two conditions are satisfied:
\begin{align}
f_k|\phi| &< T, \\
c_kf_k^2T^2 &> H^2.
\end{align}
Using the relations $|\phi|\simeq \sqrt{c}H/\sqrt{2}\lambda$ and $T \simeq (HT_{\rm RH}^2M_{\rm Pl})^{1/4}$, which hold during the matter-dominated era after inflation,
we reduce them into
\begin{align}
H &< \left(\frac{4\lambda^4T_{\rm RH}^2M_{\rm Pl}}{c^2f_k^4}\right)^{1/3}, \\
H &< (c_k^2f_k^4T_{\rm RH}^2M_{\rm Pl})^{1/3}.
\end{align}
Therefore, if the thermal mass dominates, the Hubble parameter at the beginning of the oscillation is given by
\begin{equation}
H_{\rm osc} \simeq \mathrm{min}\left[\left(\frac{4\lambda^4T_{\rm RH}^2M_{\rm Pl}}{c^2f_k^4}\right)^{1/3},\ (c_k^2f_k^4T_{\rm RH}^2M_{\rm Pl})^{1/3}\right]. \label{H_osc_therm_mass}
\end{equation}
After the oscillation, the amplitude of the AD field shifts as $|\phi|\propto R^{-3/2}T^{-1/2}\propto t^{-7/8}$~\cite{Mukaida:2012qn},
which is a bit slower than the zero-temperature case $|\phi|\propto R^{-3/2}\propto t^{-1}$.
However, we expect that the late time contribution in Eq.~\eqref{n_t_osc} becomes insignificant due to the rapidly oscillating factor $\delta_{\rm eff}$
and the amount of the baryon asymmetry is almost fixed at $H\simeq H_{\rm osc}$.

Next, consider the case where the thermal log term [Eq.~\eqref{Vphi_thermal_log}] dominates.
The oscillation of the AD field occurs when the following condition is satisfied:
\begin{equation}
a_g\alpha_g^2T^4/|\phi|^2 > H^2.
\end{equation}
From this condition, we find
\begin{equation}
H_{\rm osc} \simeq \left(\frac{2a_g\alpha_g^2\lambda^2T_{\rm RH}^2M_{\rm Pl}}{c}\right)^{1/3}. \label{H_osc_therm_log}
\end{equation}
After the oscillation, the amplitude of the AD field shifts as $|\phi|\propto R^{-3}T^{-2}\propto t^{-3/2}$~\cite{Mukaida:2012qn},
and the amount of the baryon asymmetry is almost fixed at $H\simeq H_{\rm osc}$.

Combining Eqs.~\eqref{H_osc_therm_mass} and~\eqref{H_osc_therm_log}, we obtain the general expression for the onset of the oscillation
\begin{equation}
H_{\rm osc} \simeq \mathrm{max}\left[m_{\phi},\ \mathrm{min}\left[\left(\frac{4\lambda^4T_{\rm RH}^2M_{\rm Pl}}{c^2f_k^4}\right)^{1/3},\ (c_k^2f_k^4T_{\rm RH}^2M_{\rm Pl})^{1/3}\right],\ \left(\frac{2a_g\alpha_g^2\lambda^2T_{\rm RH}^2M_{\rm Pl}}{c}\right)^{1/3} \right].
\end{equation}
From this result, we suspect that the early oscillation does not occur if $\lambda$ is sufficiently small.
It should be emphasized that the estimation of the net B asymmetry is not affected by the early oscillation due to thermal corrections,
since $n(t_{\rm osc}) \propto |\phi(t_{\rm osc})|^3/H_{\rm osc} \propto H_{\rm osc}^2$ and the ratio $n/s$ does not depend on $H_{\rm osc}$ [see Eqs.~\eqref{n_t_osc} and~\eqref{n_ov_s}].
However, due to the early oscillation, the estimation of the charge of Q-balls is modified, which affects the condition for the survival of Q-balls from evaporation.

The condition that the thermal mass term~\eqref{Vphi_thermal_mass} dominates over the thermal log term~\eqref{Vphi_thermal_log} is given by
\begin{equation}
\mathrm{min}\left[\left(\frac{4\lambda^4T_{\rm RH}^2M_{\rm Pl}}{c^2f_k^4}\right)^{1/3},\ (c_k^2f_k^4T_{\rm RH}^2M_{\rm Pl})^{1/3}\right] > \left(\frac{2a_g\alpha_g^2\lambda^2T_{\rm RH}^2M_{\rm Pl}}{c}\right)^{1/3},
\end{equation}
which can be rewritten as
\begin{equation}
7\times 10^{-12}a_g^{1/2}c^{1/2}\left(\frac{\alpha_g}{0.1}\right)\left(\frac{f_k}{10^{-5}}\right)^2 < \lambda < 4\times 10^{-10}a_g^{-1/2}c^{1/2}\left(\frac{0.1}{\alpha_g}\right)\left(\frac{c_k}{0.5}\right)\left(\frac{f_k}{10^{-5}}\right)^2. \label{osc_by_therm_mass}
\end{equation}
The early oscillation due to the thermal mass term occurs only if both of two conditions~\eqref{osc_by_therm_mass} and $H_{\rm osc} > m_{\phi}$, where $H_{\rm osc}$ is given by Eq.~\eqref{H_osc_therm_mass}, are simultaneously satisfied.
In this case, stable Q-balls are not formed since the potential is not flatter than $\phi^2$.
In particular, if the value of $f_k$ is as large as $\mathcal{O}(1)$, it occurs in the parameter region where
the sphaleron erasure effect 
is significant [i.e. the region where Eq.~\eqref{sphaleron_lambda} is not satisfied].
Such a parameter region is excluded since the primordial B/L number created due to the AD mechanism is erased until the epoch of the EWPT.

The condition that the early oscillation occurs due to the thermal log term~\eqref{Vphi_thermal_log} is given by
\begin{equation}
m_{\phi} < \left(\frac{2a_g\alpha_g^2\lambda^2T_{\rm RH}^2M_{\rm Pl}}{c}\right)^{1/3},
\end{equation}
except for the region given by Eq.~\eqref{osc_by_therm_mass}.
This corresponds to the bound
\begin{equation}
\lambda > 1\times 10^{-9}c^{1/2}a_g^{-1/2}\left(\frac{0.1}{\alpha_g}\right)\left(\frac{10^5\mathrm{GeV}}{T_{\rm RH}}\right)\left(\frac{m_{\phi}}{1\mathrm{TeV}}\right)^{3/2}. \label{osc_by_threm_log}
\end{equation}
Hence the oscillation by the thermal log term occurs for a large value of $\lambda$. 
In the following subsections, we discuss the fate of Q-balls for both cases wherein the condition~\eqref{osc_by_threm_log}
is satisfied or not.
We will see that the conditions for the survival and stability of Q-balls are not satisfied in the parameter region given by Eq.~\eqref{osc_by_threm_log}.
Combined with the fact that stable Q-balls are not formed in the region given by Eq.~\eqref{osc_by_therm_mass}, we expect that Q-balls are always destructed
once the thermal effects become relevant.

%%%%%%%%%%%%%%%%%%%%%%%%%%%%%%%%%%%%%%%%%%%%%%%%%%%%%%%%%%%%%%%%
\subsection{\label{sec4-2}Evaporation of Q-balls}
%%%%%%%%%%%%%%%%%%%%%%%%%%%%%%%%%%%%%%%%%%%%%%%%%%%%%%%%%%%%%%%%
First of all, let us consider the Q-ball solution without the finite temperature corrections.
In the framework of the gravity mediated supersymmetry breaking, the potential for the AD field including one-loop radiative corrections is given by~\cite{Enqvist:1997si}
\begin{equation}
V(\phi) = m_{\phi}^2|\phi|^2\left[1+K\ln\left(\frac{|\phi|^2}{M_*^2}\right)\right],
\end{equation}
where $K$ is a constant whose absolute value is $\mathcal{O}(0.01-0.1)$, and $M_*$ is a renormalization scale.
In the above expression, we ignored the higher order terms proportional to $\lambda$.
The potential of this form leads to the formation of Q-balls if $K$ is negative~\cite{Enqvist:1997si,Enqvist:1998en,Kasuya:2000wx,Hiramatsu:2010dx}, which we assume hereafter.
The charge of Q-balls is estimated as
\begin{align}
Q &= \frac{4}{3}\pi R_Q^3 n_B(t_{\rm form}), \\
R_Q^2 &\simeq \frac{2}{m_{\phi}^2|K|}, \label{radius_Q_zero}
\end{align}
where $R_Q$ is the radius of the Q-ball, and $n_B(t_{\rm form})$ is the charge density of the AD field at the formation time $t_{\rm form}$ of Q-balls.
Let us call this kind of Q-ball configuration ``gravity-mediation type".

To obtain the charge of Q-balls explicitly, we note that the B number at the time $t_{\rm form}$ is given by
\begin{equation}
n_B(t_{\rm form}) = \left(\frac{H_{\rm form}}{H_{\rm osc}}\right)^2 n_B(t_{\rm osc}) = \frac{\sqrt{2c^3}}{9\lambda^2}|a_m|m_{3/2}\tilde{\delta}_{\rm eff}H_{\rm form}^2,
\end{equation}
where $H_{\rm form}$ is the Hubble parameter at the time $t=t_{\rm form}$,
and we used Eq.~\eqref{n_t_osc} for $n(t_{\rm osc})$ with $\ln(t_{\rm osc}/t_{\rm inf})\simeq 1$ and $n_B(t_{\rm osc})=|n(t_{\rm osc})|/3$.
$H_{\rm form}$ can be estimated from the time scale in which the instability in the AD field grows~\cite{Enqvist:1998en},
\begin{align}
H_{\rm form} &= \frac{2m_{\phi}|K|}{\alpha}, \\
\alpha &= \ln\left(\frac{\phi(t_{\rm osc})}{\delta\phi(t_{\rm osc})}\right),
\end{align}
where $\delta\phi(t_{\rm osc})$ is the fluctuation of $\phi$ at the beginning of the oscillation $t_{\rm osc}$.
The magnitude of $\delta\phi(t_{\rm osc})$ can be related to the wavelength
$\lambda_{\rm max} \simeq 2\pi(2m_{\phi}^2|K|)^{-1/2}$ on which the instability grows with the fastest rate~\cite{Enqvist:1999mv}:
\begin{equation}
\delta\phi(t_{\rm osc}) \simeq \frac{1}{\lambda_{\rm max}} \simeq \frac{\left(2m_{\phi}^2|K|\right)^{1/2}}{2\pi}.
\end{equation}
Together with the fact that $\phi(t_{\rm osc})=\sqrt{c}H_{\rm osc}/\sqrt{2}\lambda\simeq \sqrt{c}m_{\phi}/\sqrt{2}\lambda$,
we find
\begin{equation}
\alpha = \ln\left(\frac{\sqrt{c}\pi}{\lambda |K|^{1/2}}\right).
\end{equation}
The value of $\alpha$ slightly depends on $\lambda$, but typically it takes a value of $\mathcal{O}(10)$. 
Using the ingredients obtained above, the charge of the Q-ball is estimated as
\begin{align}
Q &\simeq \frac{64\pi}{27}\frac{m_{3/2}}{m_{\phi}}\frac{|K|^{1/2}c^{3/2}|a_m|\tilde{\delta}_{\rm eff}}{\lambda^2\alpha^2} \nonumber\\
&\simeq 3\times 10^9c^{3/2}\left(\frac{m_{3/2}}{m_{\phi}}\right)\left(\frac{|K|}{0.01}\right)^{1/2}\left(\frac{10^{-6}}{\lambda}\right)^2\left(\frac{15}{\alpha}\right)^2, \label{Q_init_zero}
\end{align}
where we used $|a_m|=\tilde{\delta}_{\rm eff}=1$ in the last equality.

It is known that the charge of Q-balls diffuses away due to the coupling with the thermal bath.
Q-balls evaporate into the surrounding thermal plasma, if the chemical potential of the Q-ball $\mu_Q$ is much larger than that of the surrounding plasma $\mu_p$.
The evaporation rate is estimated as~\cite{Laine:1998rg}
\begin{equation}
\Gamma_{\rm evap} = -4\pi R_Q^2D_e(\mu_Q-\mu_p)T^2, \label{Gamma_evap}
\end{equation}
where the numerical coefficient takes a value $D_e\simeq 1$ for $T>m_{\phi}$, while the evaporation into 
sfermions is exponentially suppressed for 
$T<m_{\phi}$~\cite{Laine:1998rg,Banerjee:2000mb}\footnote{Even for $T<m_{\phi}$, it is possible to evaporate into other particles whose masses are lighter than $m_{\phi}$.
On the other hand, our purpose here is to estimate the amount of the charge transfer $\Delta Q$ above the temperature $T_{\rm erasure}$ at which the sphaleron erasure effect becomes irrelevant,
and it is enough to evaluate $\Delta Q$ up to $T\gtrsim T_{\rm erasure} \approx \mathcal{O}(m_{\tilde{f}}) \approx \mathcal{O}(m_{\phi})$.
Therefore, in this section we neglect the evaporation effect occurring at $T<m_{\phi}$.}.
However, Q-balls might achieve the chemical equilibrium with the surrounding plasma at the high temperature, and the charge transfer becomes insufficient.
In this case, the charges inside Q-balls are taken away by the diffusion effect. The rate for the charge transfer due to the diffusion is given by~\cite{Banerjee:2000mb}
\begin{equation}
\Gamma_{\rm diff} = -4\pi D_dR_Q\mu_QT^2, \label{Gamma_diff}
\end{equation}
where $D_d=a_d/T$ and the coefficient $a_d$ takes a value of 4$-$6.
The ratio between the evaporation rate~\eqref{Gamma_evap} and the diffusion rate~\eqref{Gamma_diff} becomes
\begin{equation}
\frac{\Gamma_{\rm diff}}{\Gamma_{\rm evap}} = \frac{a_d}{\sqrt{2}}|K|^{1/2}\left(\frac{m_{\phi}}{T}\right).
\end{equation}
The time scale for charge transfer is determined
by the diffusion effect ($\Gamma_{\rm diff}<\Gamma_{\rm evap}$)
for $T>T_*\equiv a_d|K|^{1/2}m_{\phi}/\sqrt{2}$.

The amount of the charge transfer
is estimated by integrating the following equation up to $T\gtrsim m_{\phi}$:
\begin{equation}
\frac{dQ}{dT} =
\left\{
\begin{array}{ll}
 \frac{dt}{dT}\Gamma_{\rm diff} & \mathrm{for}\quad T>T_*, \\
 \frac{dt}{dT}\Gamma_{\rm evap} & \mathrm{for}\quad T<T_*, 
\end{array}
\right.
\label{dQdT_zero}
\end{equation}
with $T=(HT_{\rm RH}^2M_{\rm Pl})^{1/4}$ for $T>T_{\rm RH}$ and $T=(90/\pi^2g_*(T))^{1/4}\sqrt{HM_{\rm Pl}}$ for $T<T_{\rm RH}$.
For the case with $T_{\rm RH}<m_{\phi}$, Eq.~\eqref{dQdT_zero} becomes\footnote{Here and hereafter we evaluate the amount of the charge transfer 
by assuming $T_*>m_{\phi}$.
$T_*$ can become smaller than $m_{\phi}$ depending on the values of the numerical coefficients such as $a_d$ and $K$, but in this case
the results are just modified by a factor of $\mathcal{O}(1)$, which does not affect the subsequent discussions on the baryogenesis significantly.}
\begin{equation}
\frac{dQ}{dT} \simeq 
\left\{
\begin{array}{ll}
 \frac{32\sqrt{2}\pi a_dT_{\rm RH}^2M_{\rm Pl}}{3|K|^{1/2}T^4} & \mathrm{for}\quad T>T_*, \\
\frac{64\pi T_{\rm RH}^2M_{\rm Pl}}{3|K|m_{\phi}T^3} & \mathrm{for}\quad m_{\phi}<T<T_*, 
\end{array}
\right.
 \label{dQdT_zero_1}
\end{equation}
where we used $\mu_Q\sim m_{\phi}$~\cite{Enqvist:1998en}.
Integrating Eq.~\eqref{dQdT_zero_1}, we obtain the amount of the charge 
transfer for the case with $T_{\rm RH}<m_{\phi}$:
\begin{equation}
\Delta Q \sim
\left\{
\begin{array}{ll}
 \frac{128\pi T_{\rm RH}^2M_{\rm Pl}}{9a_d^2|K|^2m_{\phi}^3} \sim 1\times 10^{16}a_d^{-2}\left(\frac{0.01}{|K|}\right)^2\left(\frac{T_{\rm RH}}{10^5\mathrm{GeV}}\right)^2\left(\frac{10^6\mathrm{GeV}}{m_{\phi}}\right)^3 & \mathrm{for}\quad T>T_*, \\
\frac{32\pi T_{\rm RH}^2M_{\rm Pl}}{3|K|m_{\phi}^3}\sim 8\times 10^{13}\left(\frac{0.01}{|K|}\right)\left(\frac{T_{\rm RH}}{10^5\mathrm{GeV}}\right)^2\left(\frac{10^6\mathrm{GeV}}{m_{\phi}}\right)^3  & \mathrm{for}\quad m_{\phi}<T<T_*.
\end{array}
\right.
\label{DeltaQ_zero_1}
\end{equation}
Note that the amount of the charge transfer becomes larger at $T>T_*$ compared with that at $m_{\phi}<T<T_*$.

Similarly, for the case with $T_{\rm RH}>m_{\phi}$, we have
\begin{equation}
\frac{dQ}{dT} = 
\left\{
\begin{array}{ll}
\frac{32\sqrt{2}\pi a_dT_{\rm RH}^2 M_{\rm Pl}}{3|K|^{1/2}T^4} & \mathrm{for}\quad T>T_{\rm RH}, \\
8a_d\left(\frac{45}{g_*|K|}\right)^{1/2}\frac{M_{\rm Pl}}{T^2} & \mathrm{for}\quad T_*<T<T_{\rm RH}, \\
8\left(\frac{90}{g_*}\right)^{1/2}\frac{M_{\rm Pl}}{|K|m_{\phi}T} & \mathrm{for}\quad m_{\phi}<T<T_*.
\end{array}
\right.
\label{dQdT_zero_2}
\end{equation}
Integrating Eq.~\eqref{dQdT_zero_2}, we find
\begin{equation}
\Delta Q \sim \left\{
\begin{array}{ll}
\frac{32\sqrt{2}\pi a_dM_{\rm Pl}}{9|K|^{1/2}T_{\rm RH}} \sim 4\times 10^{15} a_d\left(\frac{0.01}{|K|}\right)^{1/2}\left(\frac{10^5\mathrm{GeV}}{T_{\rm RH}}\right) & \mathrm{for}\quad T>T_{\rm RH}, \\
8\left(\frac{90}{g_*}\right)^{1/2}\frac{M_{\rm Pl}}{|K|m_{\phi}} \sim 2\times 10^{18}\left(\frac{100}{g_*}\right)^{1/2}\left(\frac{0.01}{|K|}\right)\left(\frac{10^3\mathrm{GeV}}{m_{\phi}}\right) & \mathrm{for}\quad T_*<T<T_{\rm RH}, \\
8\left(\frac{90}{g_*}\right)^{1/2}\frac{M_{\rm Pl}}{|K|m_{\phi}} \sim 2\times 10^{18}\left(\frac{100}{g_*}\right)^{1/2}\left(\frac{0.01}{|K|}\right)\left(\frac{10^3\mathrm{GeV}}{m_{\phi}}\right) & \mathrm{for}\quad m_{\phi}<T<T_*.
\end{array}
\right. 
\label{DeltaQ_zero_2}
\end{equation}
The amount of the charge transfer becomes larger at $m_{\phi}<T<T_{\rm RH}$ compared with that at $T>T_{\rm RH}$.

Q-balls survive from the evaporation if $Q>\Delta Q$ is satisfied.
From Eqs.~\eqref{Q_init_zero},~\eqref{DeltaQ_zero_1}, and~\eqref{DeltaQ_zero_2},
we find that this condition leads to the upper bound on the R-parity violating couplings:
\begin{align}
\lambda < 
\left\{
\begin{array}{ll}
5\times10^{-10}c^{3/4}a_d\left(\frac{|K|}{0.01}\right)^{5/4}\left(\frac{15}{\alpha}\right)\left(\frac{10^5\mathrm{GeV}}{T_{\rm RH}}\right)
\left(\frac{m_{3/2}}{m_{\phi}}\right)^{1/2}\left(\frac{m_{\phi}}{10^6\mathrm{GeV}}\right)^{3/2} & \mathrm{for} \quad T_{\rm RH}< m_{\phi}, \\
4\times10^{-11}c^{3/4}\left(\frac{|K|}{0.01}\right)^{3/4}\left(\frac{15}{\alpha}\right)\left(\frac{g_*}{100}\right)^{1/4}\left(\frac{m_{3/2}}{m_{\phi}}\right)^{1/2}\left(\frac{m_{\phi}}{10^3\mathrm{GeV}}\right)^{1/2} & \mathrm{for} \quad T_{\rm RH} > m_{\phi}.
\end{array}
\right.
\label{bound_lambda_Qevap}
\end{align}
Except for the case where $m_{\phi}$ and $T_{\rm RH}$ are extremely high, this bound lies below the sphaleron erasure bound [Eq.~\eqref{sphaleron_lambda}].
Therefore, we expect that Q-balls are not likely to preserve the B number against the sphaleron erasure effect.
We will see below that the condition becomes even worse if we consider the finite temperature effects.

If the oscillation occurs due to the thermal log term [Eq.~\eqref{Vphi_thermal_log}], the configuration of Q-balls becomes different from that of the gravity-mediation type.
This kind of Q-ball configuration is called ``thermal log type"~\cite{Kasuya:2001hg,Kasuya:2010vq}, and its radius and chemical potential are estimated as~\cite{Laine:1998rg}
\begin{equation}
R_Q \sim \frac{1}{\sqrt{2}}\frac{Q^{1/4}}{T}, \qquad \mu_Q\sim TQ^{-1/4}. \label{Q_configuration_therm_log}
\end{equation}
From the numerical simulations, the charge of Q-balls is fitted as
\begin{equation}
Q = \beta\left(\frac{|\phi(t_{\rm osc})|}{T_{\rm osc}}\right)^4, \label{Q_beta}
\end{equation}
where $T_{\rm osc}$ is the temperature of radiation at the time $t=t_{\rm osc}$, and $\beta\simeq 2\times 10^{-3}$~\cite{Kasuya:2010vq}. 
Substituting $|\phi(t_{\rm osc})|=\sqrt{c}H_{\rm osc}/\sqrt{2}\lambda$, $T_{\rm osc}\simeq (H_{\rm osc}T_{\rm RH}^2M_{\rm Pl})^{1/4}$, and Eq.~\eqref{H_osc_therm_log}
into Eq.~\eqref{Q_beta}, we find
\begin{equation}
Q \simeq 1\times 10^7c a_g\left(\frac{\alpha_g}{0.1}\right)^2\left(\frac{\beta}{2\times 10^{-3}}\right)\left(\frac{10^{-6}}{\lambda}\right)^2. \label{Q_init_therm_log}
\end{equation}
Note that the initial charge of Q-balls does not depend on the reheating temperature $T_{\rm RH}$.

As the temperature decreases, however, the correction due to the finite temperature effect becomes negligible, and the configuration of Q-balls changes
into the gravity-mediation type~\cite{Kawasaki:2006yb}.
We expect that this transformation occurs when the temperature becomes as low as 
$T^4\lesssim m_{\phi}^2\phi_c^2$, where $\phi_c$ is the value of the AD field at the center of the gravity-mediation type Q-ball.
To estimate $\phi_c$, we use the analytic expression for the charge of the Q-ball~\cite{Enqvist:1998en}
\begin{align}
Q \simeq \left(\frac{\pi}{2}\right)^{3/2}m_{\phi}\phi_c^2 R_Q^3 \simeq \left(\frac{\pi}{|K|}\right)^{3/2}\left(\frac{\phi_c}{m_{\phi}}\right)^2, \label{Q_zero_analytic}
\end{align}
where we used Eq.~\eqref{radius_Q_zero} at the last equality.
Comparing Eq.~\eqref{Q_zero_analytic} with Eq.~\eqref{Q_init_zero}, we obtain
\begin{equation}
\frac{\phi_c}{m_{\phi}} \simeq 7\times 10^2c^{3/4}\left(\frac{m_{3/2}}{m_{\phi}}\right)^{1/2}\left(\frac{|K|}{0.01}\right)\left(\frac{10^{-6}}{\lambda}\right)\left(\frac{15}{\alpha}\right). \label{phi_c_zero}
\end{equation}
Let us denote the temperature at the transformation $T_c$, for which we have
\begin{equation}
T_c \sim 3 \times 10^4\mathrm{GeV}c^{3/8}\left(\frac{|K|}{0.01}\right)^{1/2}\left(\frac{15}{\alpha}\right)^{1/2}\left(\frac{10^{-6}}{\lambda}\right)^{1/2}\left(\frac{m_{3/2}}{m_{\phi}}\right)^{1/4}\left(\frac{m_{\phi}}{10^3\mathrm{GeV}}\right).
\end{equation}
For $T<T_c$, the configuration of Q-balls is estimated as $R_Q\sim \sqrt{2}m_{\phi}^{-1}|K|^{-1/2}$ and $\mu_Q\sim m_{\phi}$, rather than Eq.~\eqref{Q_configuration_therm_log}.
We note that $m_{\phi}<T_c$ is always satisfied as long as $\lambda <\mathcal{O}(10^{-4})$.
The ratio between the evaporation rate~\eqref{Gamma_evap} and the diffusion rate~\eqref{Gamma_diff} is given by
\begin{align}
\frac{\Gamma_{\rm diff}}{\Gamma_{\rm evap}} = 
\left\{
\begin{array}{ll}
\frac{\sqrt{2}a_d}{Q^{1/4}} & \mathrm{for}\quad T>T_c, \\
\frac{a_d}{\sqrt{2}}|K|^{1/2}\left(\frac{m_{\phi}}{T}\right) & \mathrm{for}\quad m_{\phi}<T<T_c.
\end{array}
\right.
\end{align}
For $T>T_*$, we have $\Gamma_{\rm diff}<\Gamma_{\rm evap}$, and the charge transfer occurs due to the diffusion effect.

Let us estimate the amount of the charge taken away from Q-balls.
The estimation of the evaporation/diffusion rate depends on the relative size among $T_{\rm RH}$, $m_{\phi}$, and $T_c$.
First, let us consider the case with $m_{\phi}<T_c<T_{\rm RH}$. In this case, we obtain
\begin{align}
\frac{dQ}{dT} = 
\left\{
\begin{array}{ll}
\frac{32\pi a_dT_{\rm RH}^2 M_{\rm Pl}}{3\sqrt{2}T^4} & \mathrm{for}\quad T>T_{\rm RH}, \\
4a_d\left(\frac{45}{g_*}\right)^{1/2}\frac{M_{\rm Pl}}{T^2} & \mathrm{for}\quad T_c<T<T_{\rm RH}, \\
8a_d\left(\frac{45}{g_*|K|}\right)^{1/2}\frac{M_{\rm Pl}}{T^2} & \mathrm{for}\quad T_*<T<T_c, \\
8\left(\frac{90}{g_*}\right)^{1/2}\frac{M_{\rm Pl}}{|K|m_{\phi}T} & \mathrm{for}\quad m_{\phi}<T<T_*.
\end{array}
\right. \label{dQdT}
\end{align}
Integrating Eq.~\eqref{dQdT}, we obtain the charge transferred from the inside of Q-balls:
\begin{align}
\Delta Q \sim \left\{
\begin{array}{ll}
\frac{32\pi a_d M_{\rm Pl}}{9\sqrt{2}T_{\rm RH}} \sim 2\times 10^{10} a_d\left(\frac{10^9\mathrm{GeV}}{T_{\rm RH}}\right) & \mathrm{for}\quad T>T_{\rm RH}, \\
4a_d\left(\frac{45}{g_*}\right)^{1/2}\frac{M_{\rm Pl}}{T_c} \sim 6\times 10^{12}a_d\left(\frac{100}{g_*}\right)^{1/2}\left(\frac{10^6\mathrm{GeV}}{T_c}\right) & \mathrm{for}\quad T_c<T<T_{\rm RH}, \\
8\left(\frac{90}{g_*}\right)^{1/2}\frac{M_{\rm Pl}}{|K|m_{\phi}} \sim 2\times 10^{18}\left(\frac{0.01}{|K|}\right)\left(\frac{100}{g_*}\right)^{1/2}\left(\frac{10^3\mathrm{GeV}}{m_{\phi}}\right) & \mathrm{for}\quad T_*<T<T_c, \\
8\left(\frac{90}{g_*}\right)^{1/2}\frac{M_{\rm Pl}}{|K|m_{\phi}} \sim 2\times 10^{18}\left(\frac{0.01}{|K|}\right)\left(\frac{100}{g_*}\right)^{1/2}\left(\frac{10^3\mathrm{GeV}}{m_{\phi}}\right) & \mathrm{for}\quad m_{\phi}<T<T_*.
\end{array}
\right. \label{DeltaQ_1}
\end{align}
Next, let us consider the case with $m_{\phi}<T_{\rm RH}<T_c$.
In this case, we find
\begin{align}
\Delta Q \sim \left\{
\begin{array}{ll}
\frac{32\pi a_dT_{\rm RH}^2M_{\rm Pl}}{9\sqrt{2}T_c^3} \sim 2\times 10^{11} a_d\left(\frac{T_{\rm RH}}{10^5\mathrm{GeV}}\right)^2\left(\frac{10^6\mathrm{GeV}}{T_c}\right)^3 & \mathrm{for}\quad T>T_c, \\
\frac{32\sqrt{2}\pi a_d M_{\rm Pl}}{9|K|^{1/2}T_{\rm RH}} \sim 4\times 10^{15}a_d\left(\frac{0.01}{|K|}\right)^{1/2}\left(\frac{10^5\mathrm{GeV}}{T_{\rm RH}}\right) & \mathrm{for}\quad T_{\rm RH}<T<T_c, \\
8\left(\frac{90}{g_*}\right)^{1/2}\frac{M_{\rm Pl}}{|K|m_{\phi}} \sim 2\times 10^{18}\left(\frac{0.01}{|K|}\right)\left(\frac{100}{g_*}\right)^{1/2}\left(\frac{10^3\mathrm{GeV}}{m_{\phi}}\right) & \mathrm{for}\quad T_*<T<T_{\rm RH}, \\
8\left(\frac{90}{g_*}\right)^{1/2}\frac{M_{\rm Pl}}{|K|m_{\phi}} \sim 2\times 10^{18}\left(\frac{0.01}{|K|}\right)\left(\frac{100}{g_*}\right)^{1/2}\left(\frac{10^3\mathrm{GeV}}{m_{\phi}}\right) & \mathrm{for}\quad m_{\phi}<T<T_*.
\end{array}
\right. \label{DeltaQ_2}
\end{align}
Similarly, for the case with $T_{\rm RH}<m_{\phi}<T_c$, we obtain
\begin{align}
\Delta Q \sim \left\{
\begin{array}{ll}
\frac{32\pi a_dT_{\rm RH}^2M_{\rm Pl}}{9\sqrt{2}T_c^3} \sim 2\times 10^2 a_d\left(\frac{T_{\rm RH}}{10^5\mathrm{GeV}}\right)^2\left(\frac{10^9\mathrm{GeV}}{T_c}\right)^3 & \mathrm{for}\quad T>T_c, \\
\frac{128\pi T_{\rm RH}^2M_{\rm Pl}}{9a_d^2|K|^2m_{\phi}^3} \sim 1\times 10^{16}a_d^{-2}\left(\frac{0.01}{|K|}\right)^2\left(\frac{T_{\rm RH}}{10^5\mathrm{GeV}}\right)^2\left(\frac{10^6\mathrm{GeV}}{m_{\phi}}\right)^3 & \mathrm{for}\quad T_*<T<T_c, \\
\frac{32\pi T_{\rm RH}^2M_{\rm Pl}}{3|K|m_{\phi}^3} \sim 8\times 10^{13}\left(\frac{0.01}{|K|}\right)\left(\frac{T_{\rm RH}}{10^5\mathrm{GeV}}\right)^2\left(\frac{10^6\mathrm{GeV}}{m_{\phi}}\right)^3 & \mathrm{for}\quad m_{\phi}<T<T_*.
\end{array}
\right. \label{DeltaQ_3}
\end{align}

Requiring that $Q>\Delta Q$, where $Q$ is given by Eq.~\eqref{Q_init_therm_log}, we obtain the condition for the survival of Q-balls for the case
where the early oscillation occurs due to the thermal log term:
\begin{equation}
\lambda < 3\times 10^{-3}c^{1/2}a_g^{1/2}\left(\frac{\alpha_g}{0.1}\right)\left(\frac{\beta}{2\times 10^{-3}}\right)^{1/2}\Delta Q^{-1/2}, \label{Q_survival_w_early_osc}
\end{equation}
where $\Delta Q$ is given by Eq.~\eqref{DeltaQ_1},~\eqref{DeltaQ_2}, or~\eqref{DeltaQ_3}.
This bound is somewhat complicated because the value of $\Delta Q$ changes according to the relations between various parameters such as $\lambda$, $m_{\phi}$, and $T_{\rm RH}$.
In any case, it leads to a severe constraint when the early oscillation occurs due to the thermal log term.

%%%%%%%%%%%%%%%%%%%%%%%%%%%%%%%%%%%%%%%%%%%%%%%%%%%%%%%%%%%%%%%%
\subsection{\label{sec4-3}Stability of Q-balls}
%%%%%%%%%%%%%%%%%%%%%%%%%%%%%%%%%%%%%%%%%%%%%%%%%%%%%%%%%%%%%%%%
If the magnitude of the A-term at the center of the Q-ball is large, it would destabilize the Q-ball~\cite{Kawasaki:2005xc}.
To quantify the stability of Q-balls, let us define the ratio between the A-term and the mass term
\begin{equation}
\xi_Q \equiv \frac{2}{3}\frac{\lambda |a_m| m_{3/2}\phi_c}{m_{\phi}^2}, \label{xi_Q}
\end{equation}
where $\phi_c$ is the value of the AD field at the center of the Q-ball.
It was found that the instability grows with the rate $\Gamma\propto \xi_Q$, and Q-balls become unstable if this rate
exceeds the typical relaxation time scale $\sim \mathcal{O}(10)m^{-1}_{\phi}$.
The results of the numerical study~\cite{Kawasaki:2005xc} indicated that the critical value for the destabilization
is given by $\xi_{Q,\mathrm{crit}} \sim \mathcal{O}(10^{-2})$. Therefore, we expect that Q-balls are stable if the following condition
is satisfied:
\begin{equation}
\xi_Q = \frac{2}{3}\frac{\lambda |a_m| m_{3/2}\phi_c}{m_{\phi}^2} < 10^{-2}.
\end{equation}

First, let us consider the case where the finite temperature effects are negligible.
In this case, properties of the Q-ball are described by the gravity-mediation type configuration.
Substituting Eq.~\eqref{phi_c_zero} into Eq.~\eqref{xi_Q}, we find
\begin{equation}
\xi_Q \simeq 5\times 10^{-4}c^{3/4}|a_m|\left(\frac{m_{3/2}}{m_{\phi}}\right)^{3/2}\left(\frac{|K|}{0.01}\right)\left(\frac{15}{\alpha}\right).
\end{equation}
Therefore, the stability condition $\xi_Q<10^{-2}$ is satisfied for $|K|\lesssim 0.1$ and $|a_m|\lesssim 1$.
These stable Q-balls might eventually decay into light particles~\cite{Cohen:1986ct,Enqvist:1998en}.
The temperature at the time of the decay is estimated as~\cite{Kamada:2012bk}
\begin{align}
T_d & \simeq 10\mathrm{MeV}\left(\frac{m_{\phi}}{2\mathrm{TeV}}\right)^{1/2}\left(\frac{10^{28}}{Q}\right)^{1/2} \nonumber\\
& \simeq 1\times 10^7\mathrm{GeV}c^{-3/4}\left(\frac{0.01}{|K|}\right)^{1/4}\left(\frac{\alpha}{15}\right)\left(\frac{\lambda}{10^{-6}}\right)\left(\frac{m_{\phi}}{m_{3/2}}\right)^{1/2}\left(\frac{m_{\phi}}{10^3\mathrm{GeV}}\right)^{1/2},
\end{align}
where we used Eq.~\eqref{Q_init_zero} in the last equality.
Requiring that the decay occurs
below the temperature at which the sphaleron erasure effect becomes irrelevant ($T_d<T_{\rm erasure}$),
we see that the R-parity violating coupling is tightly constrained as
\begin{equation}
\lambda < 8 \times 10^{-11} c^{3/4}\left(\frac{|K|}{0.01}\right)^{1/4}\left(\frac{15}{\alpha}\right)\left(\frac{m_{3/2}}{m_{\phi}}\right)^{1/2}\left(\frac{10^3\mathrm{GeV}}{m_{\phi}}\right)^{1/2}\left(\frac{T_{\rm erasure}}{1\mathrm{TeV}}\right).
\end{equation}

The situation becomes worse if the early oscillation occurs due to the finite temperature effects.
In this case,  the value of the AD field at the center of the Q-ball is estimated as~\cite{Laine:1998rg}
\begin{equation}
\phi_c \sim \frac{1}{\sqrt{2}}TQ^{1/4}.
\end{equation}
Substituting Eq.~\eqref{Q_init_therm_log} for 
$Q$, we obtain the condition for the formation of stable Q-balls:
\begin{equation}
\xi_Q \simeq 3\times 10^{-2} c^{1/4}a_g^{1/4}|a_m|\left(\frac{\alpha_g}{0.1}\right)^{1/2}\left(\frac{\beta}{2\times 10^{-3}}\right)^{1/4}\left(\frac{\lambda}{10^{-6}}\right)^{1/2} 
\left(\frac{T}{10^6\mathrm{GeV}}\right)\left(\frac{m_{3/2}}{m_{\phi}}\right)\left(\frac{10^3\mathrm{GeV}}{m_{\phi}}\right) < 10^{-2}.\label{stability_condition_of_Q_w_early_osc}
\end{equation}
Since this bound becomes severe at high temperatures, let us estimate it at the temperature $T_{\rm form}$ of the formation of Q-balls.
To estimate $T_{\rm form}$, we note that the radius $R_{\rm form}$ of the Q-ball is comparable to the Hubble radius $H_{\rm form}^{-1}$ at the formation time:
\begin{equation}
R_{\rm form} \sim \frac{1}{\sqrt{2}}T_{\rm form}^{-1}Q^{1/4} \sim H_{\rm form}^{-1} = \frac{T_{\rm RH}^2 M_{\rm Pl}}{T_{\rm form}^4}.
\end{equation}
From this relation, we obtain
\begin{equation}
T_{\rm form} \sim 8\times 10^8\mathrm{GeV}c^{-1/12}a_g^{-1/12}\left(\frac{\alpha_g}{0.1}\right)^{-1/6}\left(\frac{\beta}{2\times 10^{-3}}\right)^{-1/12}\left(\frac{\lambda}{10^{-6}}\right)^{1/6}\left(\frac{T_{\rm RH}}{10^5\mathrm{GeV}}\right)^{2/3}. \label{T_form}
\end{equation}
Substituting Eq.~\eqref{T_form} to Eq.~\eqref{stability_condition_of_Q_w_early_osc}, we obtain
\begin{equation}
\lambda < 9\times 10^{-12}c^{-1/4}a_g^{-1/4}|a_m|^{-3/2}\left(\frac{\alpha_g}{0.1}\right)^{-1/2}\left(\frac{\beta}{2\times 10^{-3}}\right)^{-1/4}\left(\frac{T_{\rm RH}}{10^5\mathrm{GeV}}\right)^{-1}
\left(\frac{m_{\phi}}{m_{3/2}}\right)^{3/2}\left(\frac{m_{\phi}}{10^3\mathrm{GeV}}\right)^{3/2}. \label{Q_stability_lambda}
\end{equation}

%%%%%%%%%%%%%%%%%%%%%%%%%%%%%%%%%%%%%%%%%%%%%%%%%%%%%%%%%%%%%%%%
\section{\label{sec5}Baryogenesis}
%%%%%%%%%%%%%%%%%%%%%%%%%%%%%%%%%%%%%%%%%%%%%%%%%%%%%%%%%%%%%%%%
In this section, we combine the constraints obtained so far,
and discuss the parameter region where the present B asymmetry is explained.
The generated B number is given by Eq.~\eqref{n_to_s_final}, which depends on three parameters:
the magnitude of the R-parity violating coupling $\lambda$, the reheating temperature $T_{\rm RH}$,
and the gravitino mass $m_{3/2}$.
These parameters are constrained from various requirements.
In particular, we must put the bound given by Eq.~\eqref{light_element_lambda} or~\eqref{light_element_m},
since otherwise the decay of LSPs spoils the standard BBN scenario.
Also, we must take account of the sphaleron erasure bound given by Eq.~\eqref{sphaleron_lambda} or~\eqref{sphaleron_m}.
In the region where this condition is not satisfied, the dynamics of Q-balls becomes important
since there is a possibility to avoid this bound by releasing the B number from long-lived Q-balls after the 
time when the sphaleron erasure effect becomes irrelevant.
The condition for the survival of Q-balls against the evaporation is given by Eq.~\eqref{bound_lambda_Qevap} for the case where the early oscillation due to the 
thermal log term does not occur, and Eq.~\eqref{Q_survival_w_early_osc} for the case where the early oscillation occurs.
Furthermore, stable thermal log type Q-balls are not formed
if the condition given by Eq.~\eqref{Q_stability_lambda} is satisfied.
Here, the condition for the occurrence of the early oscillation is given by Eq.~\eqref{osc_by_threm_log}.

As discussed in Sec.~\ref{sec4}, we can consider two possibilities: small $\lambda$ and large $\lambda$ scenarios.
From Eq.~\eqref{n_to_s_final}, we expect that the AD mechanism works if $T_{\rm RH}$ is low for the small $\lambda$ scenario, or $T_{\rm RH}$ is high for the large $\lambda$ one.
In the following, we consider these two cases separately.

%%%%%%%%%%%%%%%%%%%%%%%%%%%%%%%%%%%%%%%%%%%%%%%%%%%%%%%%%%%%%%%%
\subsection{\label{sec5-1}Small $\lambda$ scenario}
\begin{figure}[htbp]
\centering
$\begin{array}{c}
\subfigure[]{
\includegraphics[width=0.75\textwidth]{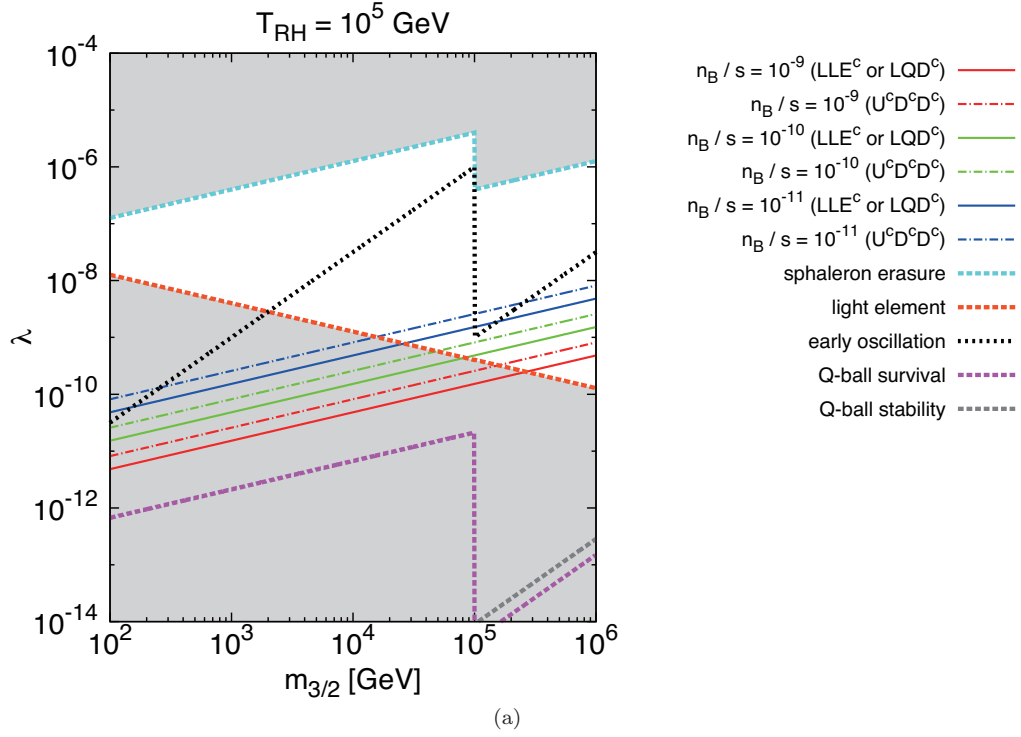}}
\\
\subfigure[]{
\includegraphics[width=0.75\textwidth]{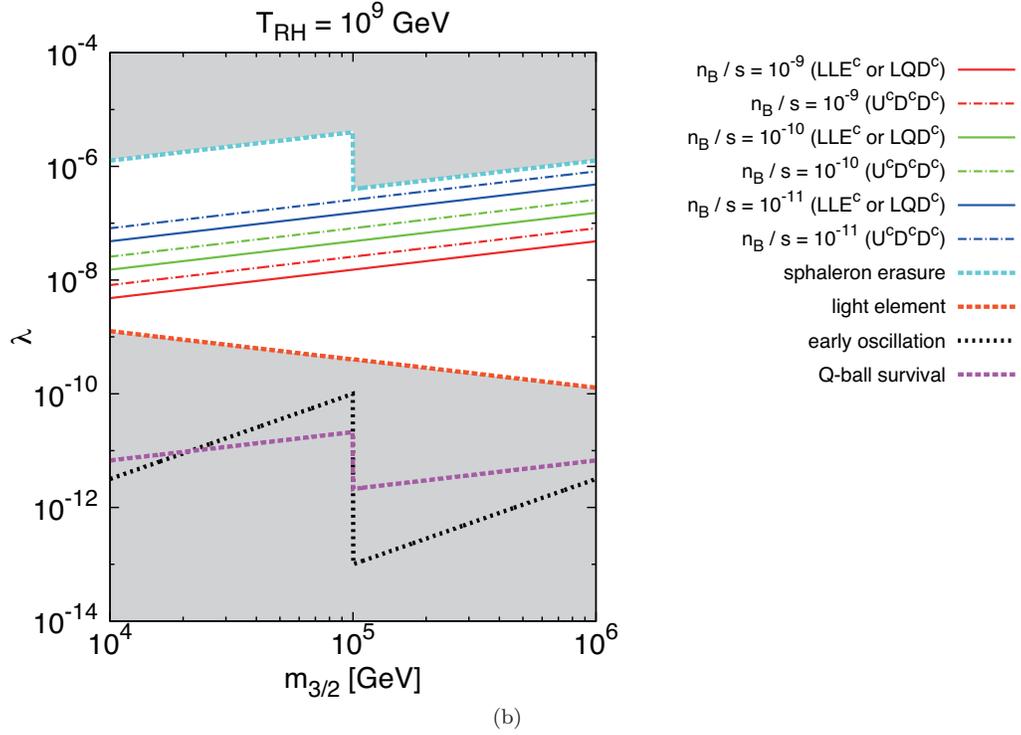}}
\end{array}$
\caption{Contour plot of $n_B/s$ as a function of $\lambda$ and $m_{3/2}$ for (a) $T_{\rm RH}=10^5\mathrm{GeV}$ and (b) $T_{\rm RH}=10^9\mathrm{GeV}$. 
The dashed cyan line ``sphaleron erasure" represents the bound [Eq.~\eqref{sphaleron_lambda}] below which the wash out effect before the EWPT is negligible.
The region below the dashed red line ``light element" [Eq.~\eqref{light_element_lambda}] is excluded 
since the unstable LSPs decay during the epoch of the BBN.
The effect of the early oscillation due to the thermal log term becomes important in the region above the dotted black line ``early oscillation", which corresponds to Eq.~\eqref{osc_by_threm_log}.
In this region, Q-balls have evaporated before the EWPT above the dashed purple line ``Q-ball survival" [Eq.~\eqref{Q_survival_w_early_osc}], and stable Q-balls are not formed above the dashed gray line ``Q-ball stability" [Eq.~\eqref{Q_stability_lambda}].
In the panel (a), the line ``Q-ball survival" [Eq.~\eqref{Q_survival_w_early_osc}] is determined by the charge transfer $\Delta Q$ in Eq.~\eqref{DeltaQ_2} with $m_{\phi}<T<T_{\rm RH}$ for $m_{3/2}<10^5\mathrm{GeV}$
and that in Eq.~\eqref{DeltaQ_3} with $T_*<T<T_c$ for $m_{3/2}>10^5\mathrm{GeV}$.
In the panel (b), it is determined by $\Delta Q$ in Eq.~\eqref{DeltaQ_2} with $m_{\phi}<T<T_{\rm RH}$.
The line of ``Q-ball stability" is not shown in panel (b) because it lies far below the range of this figure.
The region above the dashed line ``sphaleron erasure" [Eq.~\eqref{sphaleron_lambda}] is excluded since the primordial B/L number stored in Q-balls is erased until the epoch 
where equilibrium R-parity violating interactions become irrelevant.
In these plots, we take $m_{3/2}=m_{\phi}=m_{\tilde{f}}=10m_{\rm LSP}$ for $m_{3/2}<10^5\mathrm{GeV}$ with the assumption of gravity mediated supersymmetry breaking,
and $m_{3/2}=100m_{\phi}=100m_{\tilde{f}}=400m_{\rm LSP}$ for $m_{3/2}>10^5\mathrm{GeV}$ with the assumption of anomaly mediated supersymmetry breaking.}
\label{fig3}
\end{figure}

%%%%%%%%%%%%%%%%%%%%%%%%%%%%%%%%%%%%%%%%%%%%%%%%%%%%%%%%%%%%%%%%
In Fig.~\ref{fig3}, we plot the parameter dependence of the net baryon asymmetry for a fixed value of $T_{\rm RH}$.
To plot the figure, we need to specify the scenario for the supersymmetry breaking, since some conditions depend on the mass spectrum of superpartners such as $m_{\phi}$, $m_{\tilde{f}}$, and $m_{\rm LSP}$.
Here and hereafter, we assume gravity mediated supersymmetry breaking for $m_{3/2}<10^5\mathrm{GeV}$ and anomaly mediated supersymmetry breaking for $m_{3/2}>10^5\mathrm{GeV}$,
giving a hierarchy on the mass spectrum.
The specific values for relevant mass parameters are indicated in the caption of each figure.

%\if0

%%%%%%%%%%%%%%%%%%%%%%%%%%%fig3%%%%%%%%%%%%%%%%%%%%%%%%%%%%%%%%%%%
%%%%%%%%%%%%%%%%%%%%%%%%%%%%%%%%%%%%%%%%%%%%%%%%%%%%%%%%%%%%%%%%

%\fi

As shown in Fig~\ref{fig3}, we find that sphaleron erasure bound [Eq.~\eqref{sphaleron_lambda}] lies in the region where the early oscillation occurs,
and hence we must take account of the conditions for Q-balls given by Eqs.~\eqref{Q_survival_w_early_osc} and~\eqref{Q_stability_lambda}.
These two conditions turn out to be more severe than the sphaleron erasure bound,
which implies that Q-balls are likely to be destructed in the high temperature environment,
and that the primordial B/L number is washed away if the condition given by Eq.~\eqref{sphaleron_lambda} is violated.

Fortunately, we do not need to care about the survival of Q-balls if the value of $\lambda$ is smaller than the bound given by Eq.~\eqref{sphaleron_lambda},
since in such parameter regions the wash out effect becomes ineffective.
On the other hand, a small value of $\lambda$ is disfavored by the light element bound given by Eq.~\eqref{light_element_lambda}.
This situation is shown in Fig.~\ref{fig3} (a), where we take $T_{\rm RH}=10^5\mathrm{GeV}$.
The tension between sphaleron erasure bound and light element bound can be aptly avoided for the reheating temperature as high as $T_{\rm RH}=10^9\mathrm{GeV}$,
as shown in Fig.~\ref{fig3} (b). It is known that such a high reheating temperature is problematic
since the decay of gravitinos created via the scattering with particles in the thermal bath
spoils the success of the standard BBN~\cite{Khlopov:1984pf,Ellis:1984eq,Kawasaki:1994af,Kawasaki:2004yh,Kawasaki:2004qu},
but this problem can be avoided if the gravitino mass is heavier than $m_{3/2}\gtrsim \mathcal{O}(10^4-10^5)\mathrm{GeV}$.

\begin{figure}[htbp]
\centering
$\begin{array}{c}
\subfigure[]{
\includegraphics[width=0.75\textwidth]{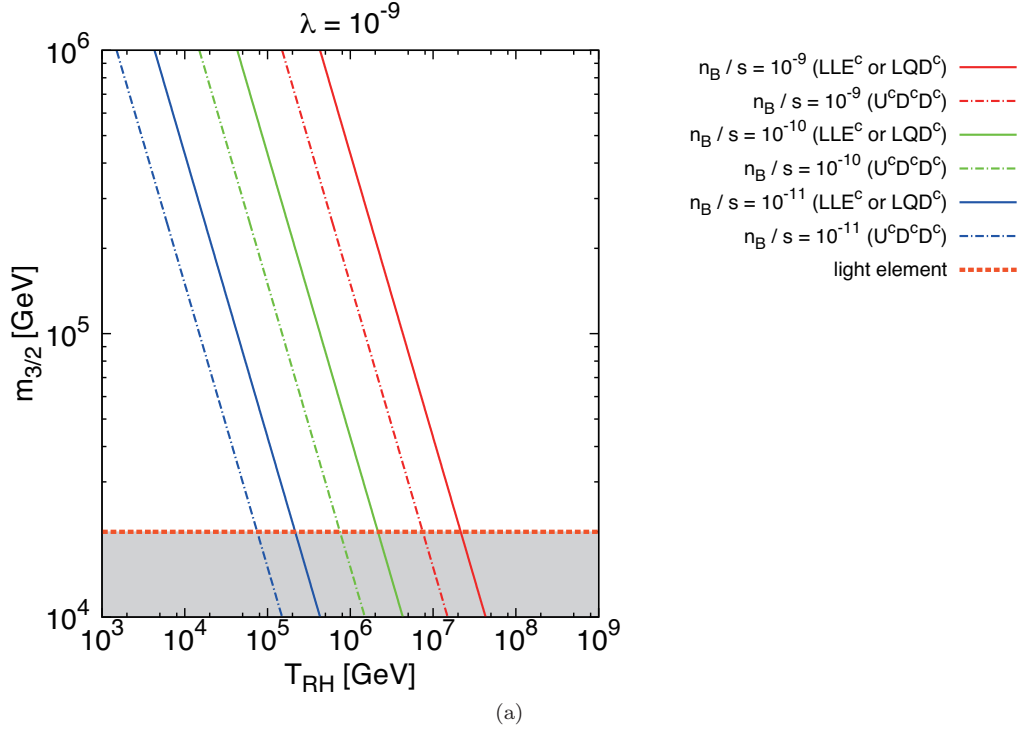}}
\\
\subfigure[]{
\includegraphics[width=0.75\textwidth]{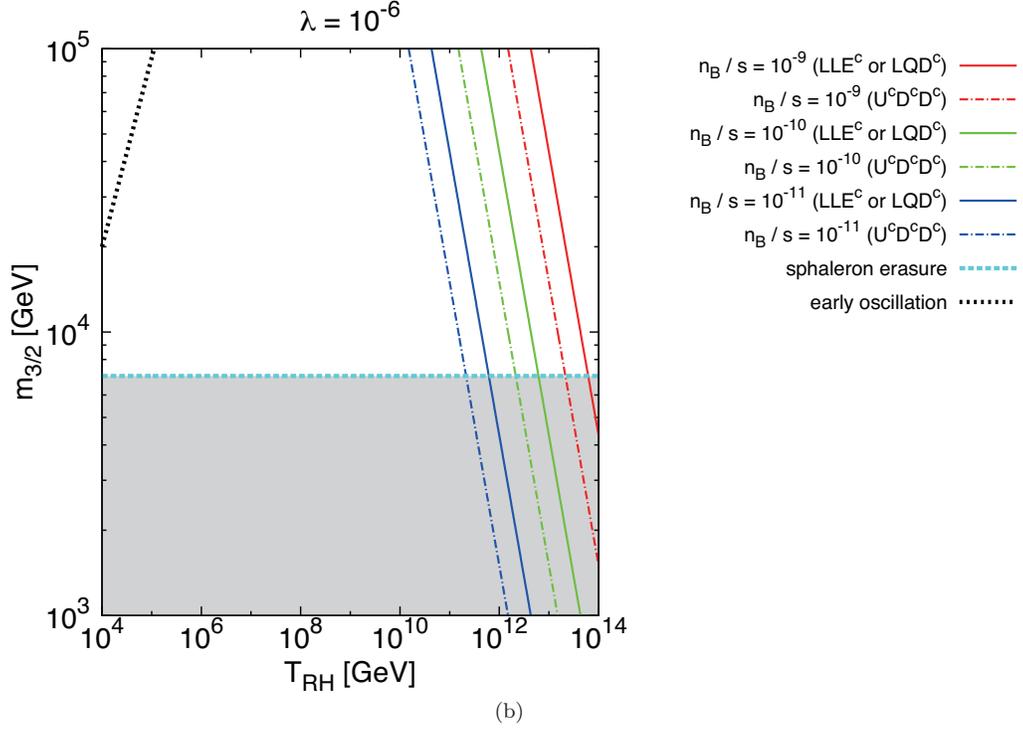}}
\end{array}$
\caption{
The same figure as Fig.~\ref{fig4} but $n_B/s$ is plotted as a function of $m_{3/2}$ and $T_{\rm RH}$ for (a) $\lambda=10^{-9}$ and (b) $\lambda=10^{-6}$. 
The dashed cyan line ``sphaleron erasure" represents the bound [Eq.~\eqref{sphaleron_m}] above which the wash out effect before the EWPT is negligible.
This ``sphaleron erasure" line is not shown in the panel (a) because it lies far below the range of the figure.
The region below the dashed red line ``light element" [Eq.~\eqref{light_element_m}] is excluded 
since the unstable LSPs decay during the epoch of the BBN.
This ``light element" line is not shown in panel (b) because it lies far below the range of the figure.
Lines corresponding to 
``Q-ball survival" [Eq.~\eqref{Q_survival_w_early_osc}] and ``Q-ball stability" [Eq.~\eqref{Q_stability_lambda}]
are not shown in panel (b), since they lie far above the range of the figure.
In these plots, we take $m_{3/2}=m_{\phi}=m_{\tilde{f}}=10m_{\rm LSP}$ with the assumption of gravity mediated supersymmetry breaking.
}
\label{fig4}
\end{figure}

In Fig.~\ref{fig4}, we also plot the parameter dependence of the net baryon asymmetry for a fixed value of $\lambda$.
From Fig.~\ref{fig4} (a), we see that for a small value of $\lambda$ ($\lambda=10^{-9}$) the successful baryogenesis ($n_B/s \approx 10^{-10}$) occurs at a low reheating temperature,
but the light element bound [Eq.~\eqref{light_element_m}] gives a lower bound on the gravitino mass.
This lower bound is replaced by the sphaleron erasure bound [Eq.~\eqref{sphaleron_m}] for a larger value of $\lambda$ ($\lambda=10^{-6}$),
as shown in Fig.~\ref{fig4} (b).
In any case, there exists a lower limit on $m_{3/2}$ for the requirement of the successful baryogenesis.

%\if0

%%%%%%%%%%%%%%%%%%%%%%%%%%%fig4%%%%%%%%%%%%%%%%%%%%%%%%%%%%%%%%%%%
%%%%%%%%%%%%%%%%%%%%%%%%%%%%%%%%%%%%%%%%%%%%%%%%%%%%%%%%%%%%%%%%

%\fi

From the results obtained above, we confirmed that the present baryon asymmetry can be explained in the broad parameter region.
Depending on the reheating temperature $T_{\rm RH}$,
the successful baryogenesis occurs in the parameter region $10^{-9}\lesssim \lambda \lesssim 10^{-6}$ with $m_{3/2}\gtrsim 10^4\mathrm{GeV}$.

%%%%%%%%%%%%%%%%%%%%%%%%%%%%%%%%%%%%%%%%%%%%%%%%%%%%%%%%%%%%%%%%
\subsection{\label{sec5-2}Large $\lambda$ scenario}
\begin{figure}[htbp]
\begin{center}
\includegraphics[width=0.75\textwidth]{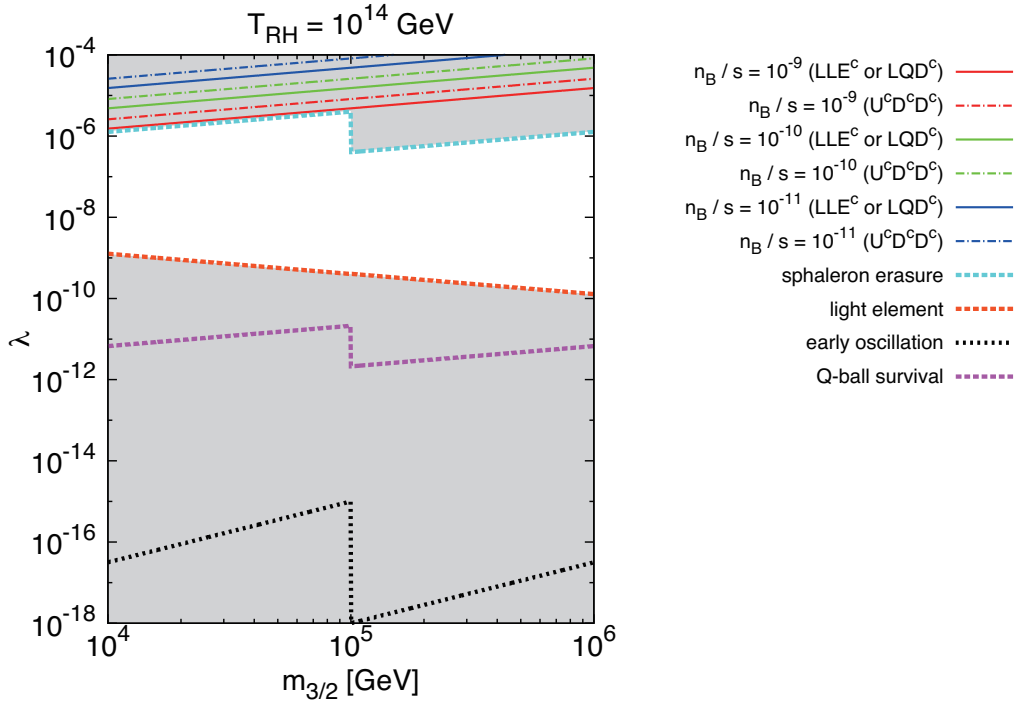}
\end{center}
\caption{
The same figure as Fig.~\ref{fig4} but $n_B/s$ is plotted as a function of $\lambda$ and $m_{3/2}$ for $T_{\rm RH}=10^{14}\mathrm{GeV}$.
The line of ``Q-ball survival" [Eq.~\eqref{Q_survival_w_early_osc}]
is determined by the charge transfer $\Delta Q$ in Eq.~\eqref{DeltaQ_1} with $m_{\phi}<T<T_c$.
The line of ``Q-ball stability" [Eq.~\eqref{Q_stability_lambda}]
is not shown because it lies far below the range of this figure.
In this plot, we take $m_{3/2}=m_{\phi}=m_{\tilde{f}}=10m_{\rm LSP}$ for $m_{3/2}<10^5\mathrm{GeV}$ with the assumption of gravity mediated supersymmetry breaking,
and $m_{3/2}=100m_{\phi}=100m_{\tilde{f}}=400m_{\rm LSP}$ for $m_{3/2}>10^5\mathrm{GeV}$ with the assumption of anomaly mediated supersymmetry breaking.
}
\label{fig5}
\end{figure}

%%%%%%%%%%%%%%%%%%%%%%%%%%%%%%%%%%%%%%%%%%%%%%%%%%%%%%%%%%%%%%%%
If the value of $\lambda$ is larger than the bound given by Eq.~\eqref{sphaleron_lambda}, the baryon asymmetry is generated only if Q-balls survive 
until the epoch where equilibrium R-parity violating interactions become irrelevant.
In this case, an adequate amount of the primordial baryon asymmetry can be generated if the reheating temperature is extremely high.
In Fig.~\ref{fig5}, we plot the parameter dependence of the net baryon asymmetry for $T_{\rm RH}= 10^{14}\mathrm{GeV}$.
We find that the effect of the early oscillation becomes important in this case, and hence the bound for the survival of Q-ball [Eq.~\eqref{Q_survival_w_early_osc}] and
that for their stability [Eq.~\eqref{Q_stability_lambda}] are applied.
As shown in Fig.~\ref{fig5}, these two bounds are below the line of the sphaleron erasure [Eq.~\eqref{sphaleron_lambda}],
and hence we conclude that above the line given by Eq.~\eqref{sphaleron_lambda}
is excluded since Q-balls have evaporated before
the sphaleron erasure effect becomes irrelevant,
and the primordial baryon asymmetry is washed away.
The leptogenesis with $L_iL_jE_k^c$ or $L_iQ_jD_k^c$ direction might be marginally allowed, but it depends on the precise values of 
factors such as $c$, $|a_m|$, and $\tilde{\delta}_{\rm eff}$ described in Sec.~\ref{sec3-2}.

It is argued that the existence of R-parity violating couplings can mitigate the constraints on the spectrum of superpartners
in the collider experiments~\cite{Carpenter:2006hs,Allanach:2012vj,Bhattacherjee:2013gr,Graham:2014vya}.
Among them, a large lepton number violating couplings ($\lambda$ and $\lambda'$) lead to the decay of gluinos into leptons,
and the null observation of such events gives a severe bound on the gluino masses.
On the other hand, the constraint on the baryon number violating couplings ($\lambda''$) is relatively weak, since the final state contains multiple jets whose identification is not straightforward
in the hadron colliders.
In that case, the magnitude of $\lambda''$ should be as large as $\mathcal{O}(10^{-5}-10^{-3})$, since otherwise superpartners do not decay inside the detectors, giving a large missing energy.
Therefore, the natural supersymmetry implies a large value of $\lambda''$ with $\lambda$ and $\lambda'$ highly suppressed.
Unfortunately, the present baryon asymmetry is not explained in this scenario,
since stable Q-balls are not formed and primordial baryon asymmetry is washed out before the epoch of the EWPT, as shown in Fig.~\ref{fig5}.

For the value as large as $\lambda\gtrsim10^{-5}$, the sphaleron erasure bound might be avoided by lowering the
reheating temperature $T_{\rm RH}< 10^{14}\mathrm{GeV}$ and
by taking an extremely large value of the gravitino mass $m_{3/2}\gtrsim10^6\mathrm{GeV}$.
In such a case we expect that the abundance of LSPs produced from the thermal bath becomes larger than the usual case, since their cross section is not fixed by
the weak scale, but its mass scale [i.e. $m_{\tilde{f}}\approx 10^{-2}m_{3/2}\gg \mathcal{O}(10^2)\mathrm{GeV}$].
Therefore, we must take account of the entropy production due to the decay of these LSPs, which is likely to reduce the efficiency of this baryogenesis scenario.
From this reason, we conclude that the present B asymmetry can hardly be explained for the case where the magnitude of the R-parity violation is as large as $\lambda\gtrsim 10^{-5}$.

%\if0

%%%%%%%%%%%%%%%%%%%%%%%%%%%fig5%%%%%%%%%%%%%%%%%%%%%%%%%%%%%%%%%%%
%%%%%%%%%%%%%%%%%%%%%%%%%%%%%%%%%%%%%%%%%%%%%%%%%%%%%%%%%%%%%%%%

%\fi

%%%%%%%%%%%%%%%%%%%%%%%%%%%%%%%%%%%%%%%%%%%%%%%%%%%%%%%%%%%%%%%%
\section{\label{sec6}Conclusion}
%%%%%%%%%%%%%%%%%%%%%%%%%%%%%%%%%%%%%%%%%%%%%%%%%%%%%%%%%%%%%%%%
In this paper, we consider a scenario where the B number of the universe is created via the AD mechanism with the R-parity
violating operators in the supersymmetric extension of the SM.
The AD mechanism works due to the presence of the trilinear R-parity violating interaction of the form given by Eq.~\eqref{W_RPV_trilinear}, which lifts the flat directions of the MSSM.
The dynamics of the AD field after inflation is calculated, and the net baryon to entropy ratio is estimated as Eq.~\eqref{n_to_s_final}.
The subsequent dynamics of Q-balls is also discussed, and it is found that Q-balls are likely to be destructed in the thermal environment.
This result leads to the difficulty in explaining the present B asymmetry with the low temperature baryogenesis due to the decay of long-lived Q-balls.
In particular, it is difficult to generate an adequate amount of the B number for the R-parity violating couplings as large as $\lambda\gtrsim 10^{-5}$.

We also find that the present B asymmetry can be explained in the broad parameter region below the bound given by Eq.~\eqref{sphaleron_lambda},
where the wash out effect of the primordial B number due to the sphaleron transition and R-parity violating interactions become negligible.
However, there is a lower bound on the R-parity violating couplings [Eq.~\eqref{light_element_lambda}] from the requirement that unstable LSPs must decay before the epoch of BBN.
To avoid these constraints, we must require the value of the R-parity violating coupling $10^{-9}\lesssim\lambda\lesssim 10^{-6}$
and the gravitino mass $m_{3/2}\gtrsim 10^4\mathrm{GeV}$.
These results imply that, for the scenario to work, the R-parity should be mildly violated and the mass scale of supersymmetry should be relatively heavy.
It will be interesting to probe such parameter regions in future experimental studies,
or discuss the origin of such a mildly broken R-parity in more fundamental theories.

%%%%%%%%%%%%%%%%%%%%%%%%%%%%%%%%%%%%%%%%%%%%%%%%%%%%%%%%%%%%%%%%
\begin{acknowledgments}
This work was supported by the Japan Society for the Promotion of Science (JSPS) Grant-in-Aid for Young Scientists (B) (No. 25800169 [TH], No. 26800121 [KN]).
K.~S.~is supported by the JSPS through research fellowships.
The work of T.~T.~is supported in part by the JSPS Grant-in-Aid for Scientific Research (No.~23740195 [TT]).
The work of M.~Y.~is supported in part by the JSPS Grant-in-Aid for Scientific Research (No.~25287054 and No.~26610062 [MY]).
\end{acknowledgments}
%%%%%%%%%%%%%%%%%%%%%%%%%%%%%%%%%%%%%%%%%%%%%%%%%%%%%%%%%%%%%%%%

\end{document}